\documentclass[12pt]{article}

\usepackage{axodraw}

\makeatletter
 
\def\paragraph{\@startsection{paragraph}{4}{\z@}{+2.00ex plus
 +1ex minus +.2ex}{1.5ex plus .2ex}{\it\normalsize}}
 
\def\section{\@startsection {section}{1}{\z@}{+3.0ex plus +1ex minus
  +.2ex}{2.3ex plus .2ex}{\normalsize\bf}}
\def\subsection{\@startsection{subsection}{2}{\z@}{+2.5ex plus +1ex
minus +.2ex}{1.5ex plus .2ex}{\normalsize\bf}}
\def\subsubsection{\@startsection{subsubsection}{3}{\z@}{+3.25ex plus
 +1ex minus +.2ex}{1.5ex plus .2ex}{\normalsize\bf}}

\@addtoreset{equation}{section}

\def\appendix{\par
 \setcounter{section}{0} 
 \setcounter{subsection}{0}
 \setcounter{equation}{0}
 \def\thesection{\Alph{section}}}
 
\expandafter\ifx\csname mathrm\endcsname\relax\def\mathrm#1{{\rm #1}}\fi
 
\makeatletter

\newcount\@tempcntc
\def\@citex[#1]#2{\if@filesw\immediate\write\@auxout{\string\citation{#2}}\fi
  \@tempcnta\z@\@tempcntb\m@ne\def\@citea{}\@cite{\@for\@citeb:=#2\do
    {\@ifundefined
       {b@\@citeb}{\@citeo\@tempcntb\m@ne\@citea
        \def\@citea{,\penalty\@m\ }{\bf ?}\@warning
       {Citation `\@citeb' on page \thepage \space undefined}}%
    {\setbox\z@\hbox{\global\@tempcntc0\csname
b@\@citeb\endcsname\relax}%
     \ifnum\@tempcntc=\z@ \@citeo\@tempcntb\m@ne
       \@citea\def\@citea{,\penalty\@m}
       \hbox{\csname b@\@citeb\endcsname}%
     \else
      \advance\@tempcntb\@ne
      \ifnum\@tempcntb=\@tempcntc
      \else\advance\@tempcntb\m@ne\@citeo
      \@tempcnta\@tempcntc\@tempcntb\@tempcntc\fi\fi}}\@citeo}{#1}}

\def\@citeo{\ifnum\@tempcnta>\@tempcntb\else\@citea
  \def\@citea{,\penalty\@m}%
  \ifnum\@tempcnta=\@tempcntb\the\@tempcnta\else
   {\advance\@tempcnta\@ne\ifnum\@tempcnta=\@tempcntb \else
\def\@citea{--}\fi
    \advance\@tempcnta\m@ne\the\@tempcnta\@citea\the\@tempcntb}\fi\fi}

\def\asymp#1%
{\mathrel{\raisebox{-.4em}{$\widetilde{\scriptstyle #1}$}}}

\def\Nequal#1%
{\mathrel{\raisebox{-.5em}{$\stackrel{=}{\scriptstyle\rm#1}$}}}
\def\dsl{\mathpalette\make@slash}
\def\make@slash#1#2{\setbox\z@\hbox{$#1#2$}%
  \hbox to 0pt{\hss$#1/$\hss\kern-\wd0}\box0}

\def\beq#1\eeq{\begin{equation}#1\end{equation}}
\def\beqar{\begin{eqnarray}}
\def\eeqar{\end{eqnarray}}
\def\barr#1{\begin{array}{#1}}
\def\earr{\end{array}}
\def\bfi{\begin{figure}}
\def\efi{\end{figure}}
\def\btab{\begin{table}}
\def\etab{\end{table}}
\def\bce{\begin{center}}
\def\ece{\end{center}}
\def\nn{\nonumber}
\def\disp{\displaystyle}

\def\al{\alpha}
\def\be{\beta}

\def\ga{\gamma}
\def\de{\delta}
\def\De{\Delta}
\def\eps{\epsilon}
\def\veps{\varepsilon}
\def\la{\lambda}

\def\si{\sigma}

\def\refeq#1{\mbox{(\ref{#1})}}

\def\reffi#1{\mbox{Fig.~\ref{#1}}}

\def\refta#1{\mbox{Table~\ref{#1}}}
\def\reftas#1{\mbox{Tables~\ref{#1}}}
\def\refse#1{\mbox{Section~\ref{#1}}}
\def\refses#1{\mbox{Sections~\ref{#1}}}
\def\refapp#1{\mbox{App.~\ref{#1}}}
\def\citere#1{\mbox{Ref.~\cite{#1}}}
\def\citeres#1{\mbox{Refs.~\cite{#1}}}

\newcommand{\GeV}{\unskip\,\mathrm{GeV}}
\newcommand{\MeV}{\unskip\,\mathrm{MeV}}

\newcommand{\pb}{\unskip\,\mathrm{pb}}

\newcommand{\ri}{{\mathrm{i}}}

\newcommand{\rd}{{\mathrm{d}}}

\newcommand{\M}{{\cal{M}}}

\def\mathswitchr#1{\relax\ifmmode{\mathrm{#1}}\else$\mathrm{#1}$\fi}

\newcommand{\PW}{\mathswitchr W}

\newcommand{\PZ}{\mathswitchr Z}

\newcommand{\Pg}{\mathswitchr g}
\newcommand{\PH}{\mathswitchr H}

\newcommand{\Pe}{\mathswitchr e}

\newcommand{\Pp}{\mathswitchr p}

\newcommand{\Pep}{\mathswitchr {e^+}}
\newcommand{\Pem}{\mathswitchr {e^-}}

\def\mathswitch#1{\relax\ifmmode#1\else$#1$\fi}

\newcommand{\Me}{\mathswitch {m_\Pe}}

\newcommand{\cut}{\mathrm{cut}}

\renewcommand{\max}{\mathrm{max}}
\newcommand{\z}{\setbox0\hbox{+}\hbox to \wd0{\hss0\hss}}

\def\slash#1{{\setbox0=\hbox{$#1$}
  \rlap{\ifdim\wd0>.7em\kern.22\wd0\else\kern.1\wd0\fi /}#1}}

\def\braket#1#2{\left\langle #1\vphantom{#2}
  \right. \kern-2.5pt\left| #2\vphantom{#1}\right\rangle }

\def\M{{\cal M}}

\def\cut{\mathswitchr{cut}}

\def\sub{{\mathrm{sub}}}
\def\sli{{\mathrm{sli}}}
\def\gsub{g^{(\sub)}}
\def\hsub{h}
\def\Gsub{G^{(\sub)}}
\def\Gsli{G^{(\sli)}}
\def\Hsub{H}
\def\cGsub{{\cal G}^{(\sub)}}
\def\cHsub{{\cal H}}
\def\cHsli{{\cal H}}
\def\bcGsub{\bar {\cal G}^{(\sub)}}
\def\bcGsli{\bar {\cal G}^{(\sli)}}
\def\bcHsub{\bar {\cal H}}

\def\draftdate{\relax}
\def\mda{\relax}
\def\mua{\relax}
\def\mla{\relax}
\def\Mda{\relax}
\def\Mua{\relax}
\def\Mla{\relax}
\def\draft{
\def\thtystars{******************************}
\def\sixtystars{\thtystars\thtystars}
\typeout{}
\typeout{\sixtystars**}
\typeout{* Draft mode!
         For final version remove \protect\draft\space in source file *}
\typeout{\sixtystars**}
\typeout{}
\def\draftdate{\today}
\def\mua{\marginpar[\boldmath\hfil$\uparrow$]%
                   {\boldmath$\uparrow$\hfil}%
                    \typeout{marginpar: $\uparrow$}\ignorespaces}
\def\mda{\marginpar[\boldmath\hfil$\downarrow$]%
                   {\boldmath$\downarrow$\hfil}%
                    \typeout{marginpar: $\downarrow$}\ignorespaces}
\def\mla{\marginpar[\boldmath\hfil$\rightarrow$]%
                   {\boldmath$\leftarrow $\hfil}%
                    \typeout{marginpar: $\leftrightarrow$}\ignorespaces}
\def\Mua{\marginpar[\boldmath\hfil$\Uparrow$]%
                   {\boldmath$\Uparrow$\hfil}%
                    \typeout{marginpar: $\uparrow$}\ignorespaces}
\def\Mda{\marginpar[\boldmath\hfil$\Downarrow$]%
                   {\boldmath$\Downarrow$\hfil}%
                    \typeout{marginpar: $\downarrow$}\ignorespaces}
\def\Mla{\marginpar[\boldmath\hfil$\Rightarrow$]%
                   {\boldmath$\Leftarrow $\hfil}%
                    \typeout{marginpar: $\leftrightarrow$}\ignorespaces}
\overfullrule 5pt
\oddsidemargin -15mm
\marginparwidth 29mm
}

\begin{document}

\thispagestyle{empty}
\def\thefootnote{\fnsymbol{footnote}}
\setcounter{footnote}{1}
\null
\strut\hfill MPP-2008-4 
\vskip 0cm
\vfill
\begin{center}
{\Large \bf 
\boldmath{Polarized QED splittings of massive fermions and
dipole subtraction for non-collinear-safe observables} \par} \vskip 2.5em
{\large
{\sc Stefan Dittmaier, Alois Kabelschacht and Tobias Kasprzik}\\[1ex]
{\normalsize \it Max-Planck-Institut f\"ur Physik (Werner-Heisenberg-Institut)\\
D-80805 M\"unchen, Germany}
}
\par \vskip 1em
\end{center} \par
\vskip 3cm 
{\bf Abstract:} \par
Building on earlier work, the dipole subtraction formalism for photonic
corrections is extended to various photon--fermion splittings
where the resulting collinear singularities lead to corrections
that are enhanced by logarithms of small fermion masses.
The difference to the earlier treatment of photon radiation is that
now no cancellation of final-state singularities is assumed, i.e.\
we allow for non-collinear-safe final-state radiation.
Moreover, we consider 
collinear fermion production from incoming photons, forward-scattering of
incoming fermions, and collinearly produced fermion--antifermion pairs.
For all cases we also provide the corresponding formulas for the
phase-space slicing approach, and
particle polarization is supported for all relevant situations.
A comparison of numerical results obtained with the proposed subtraction
procedure and the slicing method is explicitly performed for the
sample process $\Pem\ga\to\Pem\mu^-\mu^+$.
\par
\vfill
\noindent 
February 2008 \par
\vskip 1cm 
\null
\setcounter{page}{0}
\clearpage
\def\thefootnote{\arabic{footnote}}
\setcounter{footnote}{0}

\section{Introduction}

Present and future collider experiments require precise predictions
for particle reactions, i.e.\ for most of the relevant processes
radiative corrections have to be calculated.
This task becomes arbitrarily complicated if either the order
in perturbation theory (loop level) or the number of external
particles is increased, or both.
In recent years the needed techniques and concepts have received
an enormous boost from various directions; for a brief overview
we refer to some recent review articles~\cite{SMH-LH2005,DelDuca:2006wb}.

In this paper we focus on real emission corrections involving photons
at next-to-leading order (NLO). Apart from the integration over a
many-particle phase space, here the main complication is the
proper isolation of the singular parts which originate from soft
or collinear regions in phase space. To solve this problem at NLO,
two different types of methods have been developed in the past:
{\it phase-space slicing} (see, e.g., \citere{Giele:1991vf}) and 
{\it subtraction}~\cite{Ellis:1980wv,Catani:1996jh,Dittmaier:2000mb,%
Phaf:2001gc,Catani:2002hc} techniques.
In the slicing approach the singular regions are cut off from
phase space in the numerical integration and treated separately.
Employing general factorization properties of squared amplitudes
in the soft or collinear regions, the singular integrations can be
carried out analytically. In the limit of small cutoff parameters
the sum of the two contributions reproduces the full phase-space
integral. There is a trade-off between residual
cut dependences and numerical integration errors which increase
with decreasing slicing cuts; in practice, one is forced to search
for a plateau in the integrated result within some errors by varying
the slicing cut parameters.

This cumbersome procedure is not necessary within subtraction
formalisms which are based on the idea of subtracting 
a simple auxiliary function from the singular integrand and
adding this contribution again.
This auxiliary function has to be chosen in such a way that it cancels all
singularities of the original integrand so that the phase-space
integration of the difference can be performed numerically, even over the
singular regions of the original integrand.
In this difference the original matrix element can be
evaluated without regulators for soft or collinear singularities.
The auxiliary function has to be simple enough so that it can
be integrated over the singular regions analytically with the help of
regulators, when the subtracted contribution is added again. 
This singular analytical integration can be done once and for all
in a process-independent way because of the general factorization
properties of squared amplitudes in the singular regions.
At NLO several subtraction variants have been proposed in the literature~%
\cite{Ellis:1980wv,Catani:1996jh,Dittmaier:2000mb,%
Phaf:2001gc,Catani:2002hc},
some of which are quite general; at next-to-next-to-leading
order subtraction formalisms are still under construction
\cite{Weinzierl:2003fx}.

The {\it dipole subtraction formalism} certainly represents the most
frequently used subtraction technique in NLO
calculations. It was first proposed within massless QCD by
Catani and Seymour~\cite{Catani:1996jh} and subsequently generalized
to photon emission off massive fermions~\cite{Dittmaier:2000mb}%
\footnote{The case of light fermions, where masses appear as
regulators, has also been worked out in \citere{Roth:1999kk}.}
and to QCD with massive quarks~\cite{Phaf:2001gc,Catani:2002hc}.
Among the numerous applications of dipole subtraction, we merely
mention the treatment of the electroweak corrections to
$\Pep\Pem\to4\,$fermions~\cite{Denner:2005es}, which was
the first complete treatment of a $2\to4$ particle process at NLO.
The formulation~\cite{Catani:1996jh,Phaf:2001gc,Catani:2002hc}
of dipole subtraction for NLO QCD corrections assumes so-called
infrared safety of observables, i.e.\ that all soft or collinear
singularities cancel against their counterparts from the virtual 
corrections, either after parton-density redefinitions
for initial-state singularities or due to the inclusiveness of
event selection criteria in soft or collinear configurations
for final-state singularities.
In \citere{Dittmaier:2000mb} the collinear singularities from photon
radiation are regularized by physical fermion masses, and only for
final-state radiation these were assumed to cancel due to inclusiveness
of the observable. 

\begin{sloppypar}
In the following we generalize the method of \citere{Dittmaier:2000mb} 
by dropping the latter assumption and by considering also other
collinear-singular configurations involving photons, which
are regularized by physical fermion masses:
\begin{enumerate}
\item
In \refse{se:fsr} we deal with {\it non-collinear-safe final-state radiation
off light \mbox{(anti-)}fermions $f$},
where collinear singularities arise from the splitting $f^*\to f\ga$.
Here and in the following, asterisks indicate off-shell particles.
By non-collinear-safe radiation we mean that a collinear fermion--photon
system is not necessarily treated as one quasi-particle, which by
contrast is the
case in any collinear-safe observable. In collinear-safe situations,
which are usually enforced by photon recombination or a jet algorithm,
singularities from final-state radiation cancel according to the
well-known Kinoshita--Lee--Nauenberg (KLN) theorem~\cite{KLN}.
Non-collinear-safe final-state radiation off a fermion $f$, in general,
leads to corrections $\propto\alpha\ln m_f$ that are enhanced by
a logarithm of a small fermion mass $m_f$.
\item
Section~\ref{se:affst} is devoted to {\it processes with incoming photons
and outgoing light fermions}. Here the collinear-singular splitting is
$\ga\to f\bar f^*$, i.e.\ if an outgoing \mbox{(anti-)}fermion $f$ is allowed
to be scattered into the direction of the incoming photon, the cross
section receives an enhancement $\propto \ln m_f$ from this phase-space
region.
\item
In Section~\ref{se:astff} we treat {\it processes with light
fermion--antifermion pairs in the final state}, i.e.\
when an outgoing photon with low virtuality splits into an $f\bar f$ pair,
$\ga^*\to f\bar f$.
If the collinearly produced $f \bar f$ pair can be distinguished from 
a plainly emitted photon (that has not split), the considered cross
section again receives an enhancement $\propto \ln m_f$.
\item
Finally, in \refse{se:ffast} we concentrate on {\it processes with
forward-scattered light \mbox{(anti-)}fermions}, where the splitting
$f\to f\ga^*$ leads to a collinear singularity if the emitted photon
is almost real. Again this phase-space region enhances the cross section
by a factor $\propto \ln m_f$.
\end{enumerate}
While \refse{se:fsr} builds on the conventions and results of 
\citere{Dittmaier:2000mb}, \refses{se:affst}, \ref{se:astff}, and
\ref{se:ffast} are self-contained and can be read independently.
\end{sloppypar}

Of course, the considered situations could all be treated by fully
including a non-zero fermion mass $m_f$ in the calculation.
However, if $m_f$ is small compared to typical scales in the process,
which is the case for electrons or muons in almost all present and
future high-energy collider experiments, such a procedure is very
inconvenient. The presence of very small or large scale ratios jeopardizes
the numerical stability of phase-space integrations, and mass
terms significantly slow down the evaluation of matrix elements.
The subtraction technique described in the following
avoids these problems by completely isolating all
mass singularities from squared matrix elements, so that finally only
amplitudes for a massless fermion $f$ are needed.
We support particle polarization whenever relevant, in particular
for all incoming particles. In order to facilitate cross-checks in
applications, the corresponding formulas for the
phase-space slicing approach are also provided.

In \refse{se:appl} we demonstrate the use and the performance 
of the methods presented in \refses{se:affst}, \ref{se:astff}, and
\ref{se:ffast} in the example $\Pem\ga\to\Pem\mu^-\mu^+$.
A summary is given in \refse{se:summary}, and the appendices provide
more details on and generalizations of the formulas presented in the
main text. In particular, the derivation of the factorization 
formulas for processes with incoming polarized photons splitting into
light fermions and for the forward scattering of incoming polarized 
light fermions is described there.

\section{Non-collinear-safe photon radiation off final-state fermions}
\label{se:fsr}

\subsection{Dipole subtraction and non-collinear-safe observables}
\label{se:noncollsafe}

For any subtraction formalism the schematic form
of the subtraction procedure to integrate the squared matrix
element $\sum_{\lambda_\gamma} |\M_1|^2$ 
(summed over photon polarizations $\la_\ga$)
for real photon radiation
over the $(N+1)$-particle phase space $\rd\Phi_1$ reads
\beq
\int\rd\Phi_1\, \sum_{\lambda_\gamma} |\M_1|^2 =
\int\rd\Phi_1\, \Biggl(\sum_{\lambda_\gamma}|\M_1|^2-|\M_\sub|^2\Biggr)
+ \int\rd\tilde\Phi_0\,\otimes\left(\int [\rd k] \, |\M_\sub|^2\right),
\label{eq:sub}
\eeq
where $\rd\tilde\Phi_0$ is a phase-space element of the corresponding
non-radiative process and $[\rd k]$ includes the photonic phase space
that leads to the soft and collinear singularities.
The two contributions involving the subtraction function $|\M_\sub|^2$
have to cancel each other, however, they will be evaluated separately.
The subtraction function is constructed in such a way that
the difference $\sum_{\lambda_\gamma}|\M_1|^2-|\M_\sub|^2$
can be safely integrated over $\rd\Phi_1$ numerically and that the 
singular integration of $|\M_\sub|^2$ over $[\rd k]$ can be carried out
analytically, followed by a safe numerical integration over $\rd\tilde\Phi_0$.

In the dipole subtraction formalism for photon radiation, 
the subtraction function is given by \cite{Dittmaier:2000mb}
\beq
|\M_{\sub}(\Phi_1;\kappa_f)|^2 =
-\sum_{f\neq f'} Q_f \sigma_f Q_{f'} \sigma_{f'} e^2
\gsub_{ff',\tau}(p_f,p_{f'},k)
\left|\M_0\left(\tilde\Phi_{0,ff'};\tau\kappa_f\right)\right|^2,
\label{eq:m2sub}
\eeq
where the sum runs over all emitter--spectator pairs $ff'$, which are called
dipoles. 
For a final-state emitter (final-state radiation), the two possible
dipoles are illustrated in \reffi{fig:fsr}.
\bfi
\centerline{
\begin{picture}(310,110)(0,0)
\put(0,-10){
  \begin{picture}(100,120)(0,0)
  \Line(20,50)(80, 90)
  \Line(20,50)(60, 20)
  \LongArrow(30,30)(45,18)
  \LongArrow(55,85)(70,95)
  \LongArrow(55,55)(70,45)
  \Photon(50,70)(80,50){2}{4}
  \Vertex(50,70){2.5}
  \GCirc(20,50){10}{1}
  \put(85,88){$i$}
  \put(85,46){$\gamma$}
  \put(65,15){$j$}
  \put(30,14){$p_j$}
  \put(55,99){$p_i$}
  \put(56,40){$k$}
  \end{picture} } 
\put(130,-10){
  \begin{picture}(160,120)(0,0)
  \Line(80,50)(140, 90)
  \Line(80,50)( 40, 20)
  \LongArrow( 52,18)( 67, 30)
  \LongArrow(115,85)(130, 95)
  \LongArrow(115,55)(130, 45)
  \Photon(110,70)(140,50){2}{4}
  \Vertex(110,70){2.5}
  \GCirc(80,50){10}{1}
  \put(145, 88){$i$}
  \put(145, 46){$\gamma$}
  \put( 28, 18){$a$}
  \put( 65, 18){$p_a$}
  \put(115, 99){$p_i$}
  \put(116, 40){$k$}
  \end{picture} } 
\end{picture} } 
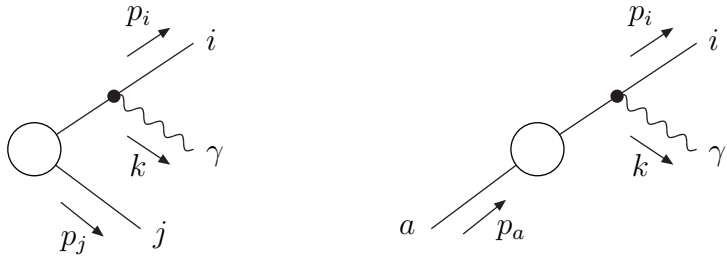
\caption{Generic diagrams for photonic final-state radiation off an
emitter $i$ with a spectator $j$ or $a$ in the final or initial state, 
respectively.}
\label{fig:fsr}
\efi
The relative charges are denoted $Q_f$, $Q_{f'}$, and
the sign factors $\si_f,\si_{f'}=\pm1$ correspond to the charge flow
($\sigma_f=+1$ for incoming fermions and outgoing antifermions, 
$\sigma_f=-1$ for outgoing fermions and incoming antifermions).
The implicitly assumed
summation over $\tau=\pm$ accounts for a possible flip in the
helicity of the emitter $f$, where $\kappa_f=\pm$ is the sign of the
helicity of $f$ both in $|\M_1|^2$ and $|\M_\sub|^2$.
The singular behaviour of the subtraction function is contained in
the radiator functions $\gsub_{ff',\tau}(p_f,p_{f'},k)$, which depend
on the emitter, spectator, and photon momenta $p_f$, $p_{f'}$, and $k$,
respectively. The squared lowest-order matrix element $|\M_0|^2$ 
of the corresponding non-radiative process enters the subtraction function
with modified emitter and spectator momenta $\tilde p_f^{(ff')}$ and
$\tilde p_{f'}^{(ff')}$.
For a final-state emitter $f$, the momenta are related by
$p_f+k\pm p_{f'}=\tilde p_f^{(ff')} \pm \tilde p_{f'}^{(ff')}$, where
$\pm$ refers to a spectator $f'$ in the final or initial state,
and the same set $\{k_n\}$ of remaining particle momenta enters 
$|\M_1|^2$ and $|\M_0|^2$.
The modified momenta are constructed in such a way that
$\tilde p_f^{(ff')} \to p_f+k$ in the collinear limit $(p_f k\to0)$.

Note that no collinear singularity exists for truly massive radiating
particles $f$, because the invariant $p_f k$ does not tend to zero
if the photon emission angle becomes small (for fixed photon energy $k^0$).
In such cases the corresponding masses are kept non-zero in all
amplitudes, in the subtraction functions, and in the kinematics,
and the subtraction procedure works without problems.
Collinear (or mass) singularities result if the mass $m_f$ of a
radiating particle is much smaller than the typical scale in the process
under consideration. In such cases it is desirable to set $m_f$ to
zero whenever possible. In a subtraction technique this means that
$m_f=0$ can be consistently used in the integral 
$\int\rd\Phi_1\, \left(\sum_{\lambda_\gamma}|\M_1|^2-|\M_\sub|^2\right)$,
but that the readded contribution $\int [\rd k] \, |\M_\sub|^2$ 
contains mass-singular terms of the form $\alpha\ln m_f$.
If such mass singularities from collinear photon radiation 
do not completely cancel against their counterparts in
the virtual corrections,
the corresponding observable is {\it not collinear safe}.
The dipole subtraction formalism as described in \citere{Dittmaier:2000mb}
is formulated to cover possible mass singularities from initial-state
radiation, but assumes collinear safety w.r.t.\ final-state radiation. 

In {\it collinear-safe} observables (w.r.t.\ final-state radiation), 
and only those are considered for light fermions
in \citere{Dittmaier:2000mb}, a collinear fermion--photon system is
treated as one quasi-particle, i.e., in the limit where $f$ and $\gamma$ 
become collinear only the sum $p_f+k$ enters the procedures of implementing 
phase-space selection cuts or of sorting an event into a histogram bin
of a differential distribution. Technically this level of inclusiveness
is reached by {\it photon recombination}, a procedure that assigns the
photon to the nearest charged particle if it is close enough to it.
Of course, different variants for such an algorithm are possible,
similar to jet algorithms in QCD.
The recombination guarantees that for each photon radiation cone
around a charged particle $f$ the energy fraction 
\beq
z_f=\frac{p_f^0}{p_f^0+k^0}
\eeq
is fully integrated over. According to the KLN theorem,
no mass singularity connected with final-state radiation remains.
Collinear safety facilitates the actual application of the subtraction
procedure as indicated in Eq.~\refeq{eq:sub}. In this case the
events resulting from the contributions of $|\M_\sub|^2$ can be 
consistently regarded as $N$-particle final states of the
non-radiative process with particle momenta as going into 
$\left|\M_0\left(\tilde\Phi_{0,ff'}\right)\right|^2$, i.e.\
the emitter and spectator momenta are given by $\tilde p_f^{(ff')}$,
$\tilde p_{f'}^{(ff')}$, respectively.
Owing to $\tilde p_f^{(ff')} \to p_f+k$ in the collinear limits, the difference
$\sum_{\lambda_\gamma}|\M_1|^2-|\M_\sub|^2$ can be integrated
over all collinear regions, because all events that differ only in the value
of $z_f$ enter cuts or histograms in the same way.
The implicit {\it full} integration over all $z_f$ in the collinear cones,
on the other hand, implies that in the analytical integration of
$|\M_\sub|^2$ over $[\rd k]$ the $z_f$ integrations can be carried out
over the whole $z_f$ range.

In {\it non-collinear-safe} observables (w.r.t.\ final-state radiation), 
not all photons within arbitrarily narrow collinear cones around outgoing
charged particles are treated inclusively. For a fixed cone axis the
integration over the corresponding variable $z_f$ is constrained by
a phase-space cut or by the boundary of a histogram bin. Consequently,
mass-singular contributions of the form $\alpha\ln m_f$ remain
in the integral.
Technically this means that the information on the variables $z_f$
has to be exploited in the subtraction procedure of Eq.~\refeq{eq:sub}.
The variables that take over the role of $z_f$ in the individual
dipole contributions in $|\M_\sub|^2$ are called $z_{ij}$ and
$z_{ia}$ in \citere{Dittmaier:2000mb}, where 
$f=i$ is a final-state emitter and $j/a$ a final-/initial-state spectator.
In the collinear limit they behave as $z_{ij}\to z_i$ and $z_{ia}\to z_i$.
Thus, the integral 
$\int\rd\Phi_1\, \left(\sum_{\lambda_\gamma}|\M_1|^2-|\M_\sub|^2\right)$ 
can be performed over the whole phase space if the events associated with
$|\M_\sub|^2$ are treated as $(N+1)$-particle event with momenta
$p_f\to z_{ff'} \tilde p_f^{(ff')}$,
$p_{f'}\to \tilde p_{f'}^{(ff')}$, and $k\to (1-z_{ff'})\tilde p_f^{(ff')}$.
This can be formalized by introducing 
a step function $\Theta_{\cut}(p_f,k,p_{f'},\{k_n\})$
on the $(N+1)$-particle phase space which is 1 if the event passes the
cuts and 0 otherwise. The set $\{k_n\}$ simply contains the momenta
of the remaining particles in the process.
Making the dependence on $\Theta_{\cut}$ explicit,
the first term on the r.h.s.\ of Eq.~\refeq{eq:sub} reads
\beqar
&& \int\rd\Phi_1\, \Biggl[
\sum_{\lambda_\gamma}|\M_1|^2
\Theta_{\cut}(p_f,k,p_{f'},\{k_n\})
\nn\\
&& \hspace{3.5em} {}
-\sum_{f\ne f'} |\M_{\sub,ff'}|^2
\Theta_{\cut}\Bigl(z_{ff'} \tilde p_f^{(ff')},(1-z_{ff'})\tilde p_f^{(ff')},
\tilde p_{f'}^{(ff')},\{k_n\}\Bigr)
\Biggr],
\eeqar
where we have decomposed the subtraction function $|\M_\sub|^2$ into
its subcontributions $|\M_{\sub,ff'}|^2$ of specific emitter--spectator
pairs $ff'$.
Apart from this refinement of the cut prescription in the subtraction 
part for non-collinear-safe observables, no modification in
$|\M_\sub|^2$ is needed. Since its construction exactly proceeds as
described in Sections~3 and 4 of \citere{Dittmaier:2000mb}, we do not 
repeat the individual steps in this paper.

However, the modification of the cut procedure
requires a generalization of the evaluation
of the second subtraction term on the r.h.s.\ of Eq.~\refeq{eq:sub},
because now the integral over $z_{ff'}$ implicitly contained in
$[\rd k]$ depends on the cuts that define the observable. 
In the following two sections
we work out the form of the necessary modifications,
where we set up the formalism in such a way that it reduces to the 
procedure described in \citere{Dittmaier:2000mb} for a collinear-safe
situation, while the non-collinear-safe case is covered upon including
extra contributions. 

\subsection{Final-state emitter and final-state spectator}
\label{se:fsr-fin}

For a final-state emitter $i$ and a final-state spectator $j$ with
masses $m_i$ and $m_j$ the integral of $\gsub_{ij,\tau}(p_i,p_j,k)$
over $[\rd k]$ is proportional to 
\beq
\Gsub_{ij,\tau}(P_{ij}^2) = \frac{\bar P_{ij}^4}{2\sqrt{\lambda_{ij}}}
\int_{y_1}^{y_2} \rd y_{ij}\, (1-y_{ij})
\int_{z_1(y_{ij})}^{z_2(y_{ij})} \rd z_{ij} \,
\gsub_{ij,\tau}(p_i,p_j,k),
\eeq
where the definitions of Sections~3.1 and 4.1 of 
\citere{Dittmaier:2000mb} are used.
There the results for $\Gsub_{ij,\tau}(P_{ij}^2)$ with generic or light masses
are given in Eqs.~(4.10) and (3.7), respectively.
In order to leave the integration over $z_{ij}$ open, the order of the two
integrations has to be interchanged, and the integral solely taken over
$y_{ij}$ is needed. Therefore, we define
\beq
\bcGsub_{ij,\tau}(P_{ij}^2,z_{ij}) = 
\frac{\bar P_{ij}^4}{2\sqrt{\lambda_{ij}}}
\int_{y_1(z_{ij})}^{y_2(z_{ij})} \rd y_{ij}\, (1-y_{ij})\,
\gsub_{ij,\tau}(p_i,p_j,k).
\label{eq:bcGsubij}
\eeq
Note that no finite photon mass $m_\ga$ is needed
in the function $\bcGsub_{ij,\tau}(P_{ij}^2,z)$ in practice,
because the soft singularity appearing
at $z\to1$ can be split off by employing a $[\dots]_+$ prescription
in the variable $z$,
\beq
\bcGsub_{ij,\tau}(P_{ij}^2,z) =
\Gsub_{ij,\tau}(P_{ij}^2) \de(1-z) 
+ \left[\bcGsub_{ij,\tau}(P_{ij}^2,z)\right]_+.
\eeq
This procedure shifts the soft singularity into the quantity
$\Gsub_{ij,\tau}(P_{ij}^2)$, which is already known from
\citere{Dittmaier:2000mb}. Moreover, the generalization to
non-collinear-safe integrals simply reduces to the extra term
$\left[\bcGsub_{ij,\tau}(P_{ij}^2,z)\right]_+$, which cancels out
for collinear-safe integrals where the full $z$-integration is carried
out.

For arbitrary values of $m_i$ and $m_j$ a compact analytical result of 
$\bcGsub_{ij,\tau}(P_{ij}^2,z)$ cannot be achieved
because of the complicated structure of the integration boundary.
Note, however, that only the limit $m_i\to0$ of a light emitter is
relevant, since for truly massive emitters no mass singularity
results. The case of a massive spectator $j$ is presented in 
\refapp{app:fsr};
here we restrict ourselves to the simpler but important special
case $m_j=0$.

In the limit $m_i\to0$ and $m_j=m_\ga=0$ the boundary of the $y_{ij}$ 
integration is asymptotically given by
\beq
y_1(z) = \frac{m_i^2(1-z)}{P_{ij}^2 z}, \quad 
y_2(z) = 1, 
\label{eq:yboundary}
\eeq
and the functions and quantities
relevant in the integrand $\gsub_{ij,\tau}$ behave as
\beq
p_i k = \frac{P_{ij}^2}{2} y_{ij}, \quad
R_{ij}(y) = 1-y, \quad r_{ij}(y) = 1.
\eeq
The evaluation of Eq.~\refeq{eq:bcGsubij} becomes very simple and yields
\beqar
\bcGsub_{ij,+}(P_{ij}^2,z) &=&
P_{ff}(z)\,\biggl[ \ln\biggl(\frac{P_{ij}^2 z}{m_i^2}\biggr)-1\biggl]
+(1+z)\ln(1-z), \nn\\ 
\bcGsub_{ij,-}(P_{ij}^2,z) &=& 1-z,
\label{eq:bcGsubij2}
\eeqar
where $P_{ff}(z)$ is the splitting function,
\beq
P_{ff}(z) = \frac{1+z^2}{1-z}.
\eeq
Equation~\refeq{eq:bcGsubij2} 
is correct up to terms suppressed by factors of $m_i$.
For completeness, we repeat the form of the full integral
$\Gsub_{ij,\tau}(P_{ij}^2)$ in the case of light masses,
\beq
\Gsub_{ij,+}(P_{ij}^2) = {\cal L}(P_{ij}^2,m_i^2) - \frac{\pi^2}{3} + 1,
\qquad
\Gsub_{ij,-}(P_{ij}^2) = \frac{1}{2},
\label{eq:Gij0}
\eeq
with the auxiliary function
\beq
{\cal L}(P^2,m^2) =
\ln\biggl(\frac{m^2}{P^2}\biggr)
\ln\biggl(\frac{m_\gamma^2}{P^2}\biggr)
+ \ln\biggl(\frac{m_\gamma^2}{P^2}\biggr)
- \frac{1}{2}\ln^2\biggl(\frac{m^2}{P^2}\biggr)
+ \frac{1}{2}\ln\biggl(\frac{m^2}{P^2}\biggr),
\label{eq:L}
\eeq
which are taken from Eqs.~(3.7) and (3.8) of \citere{Dittmaier:2000mb}.%
\footnote{If dimensional regularization is used to regularize the soft
singularity instead of a finite photon mass, the photon-mass logarithm
in ${\cal L}$ has to be replaced according to $\ln(m_\ga^2) \to
(4\pi\mu^2)^\epsilon\Gamma(1+\epsilon)/\epsilon + {\cal O}(\epsilon)$,
where $D=4-2\epsilon$ is the dimension and $\mu$ the reference mass of
dimensional regularization.}

Finally, we give the explicit form of the $ij$ contribution
$|\M_{\sub,ij}(\Phi_1)|^2$ to the phase-space integral of the 
subtraction function,
\beqar
\int\rd\Phi_1\,|\M_{\sub,ij}(\Phi_1;\kappa_i)|^2
&=& -\frac{\alpha}{2\pi} Q_i\sigma_i Q_j\sigma_j  \,
\int\rd\tilde\Phi_{0,ij}\,
\int_0^1 \rd z\, 
\nn\\[.3em]
&& {}\times
\left\{ \Gsub_{ij,\tau}(P_{ij}^2) \de(1-z) 
+ \left[\bcGsub_{ij,\tau}(P_{ij}^2,z)\right]_+ \right\}
\\[.3em]
&& {}\times
|\M_0(\tilde p_i,\tilde p_j;\tau\kappa_i)|^2 \,
\Theta_{\cut}\Bigl(p_i=z\tilde p_i,k=(1-z)\tilde p_i,\tilde p_j,\{k_n\}\Bigr),
\nn
\label{eq:zintij}
\eeqar
generalizing Eq.~(3.6) of \citere{Dittmaier:2000mb}.
While $\tilde p_i,\tilde p_j,\{k_n\}$ are the momenta corresponding to the
generated phase-space point in $\tilde\Phi_{0,ij}$, the momenta
$p_i$ and $k$ result from $\tilde p_i$ via a simple rescaling with the 
independently generated variable $z$. The invariant $P_{ij}^2$ is
calculated via $P_{ij}^2=(\tilde p_i+\tilde p_j)^2$ independently of $z$.
The arguments of the step function $\Theta_{\cut}(p_i,k,\tilde p_j,\{k_n\})$
indicate on which momenta phase-space cuts are imposed.

\begin{sloppypar}
For unpolarized fermions the results of this section have already 
been described in \citere{Bredenstein:2005zk}, where electroweak
radiative corrections to the processes $\ga\ga\to\PW\PW\to4\,$fermions
were calculated. In this calculation the results for non-collinear-safe
differential cross sections were also cross-checked against results obtained
with phase-space slicing.
Another comparison between the described subtraction procedure and 
phase-space slicing has been performed in the calculation of
electroweak corrections to the Higgs decay processes
$\PH\to\PW\PW/\PZ\PZ\to4\,$fermions \cite{Bredenstein:2006rh}.
\end{sloppypar}

\subsection{Final-state emitter and initial-state spectator}
\label{se:fsr-ini}

For the treatment of a final-state emitter $i$ and an
initial-state spectator $a$, we consistently make use of the 
definitions of Sections~3.2 and 4.2 of \citere{Dittmaier:2000mb}.
In this paper we only consider light particles in the initial state, 
because the masses of incoming particles are much smaller than the
scattering energies at almost all present and future colliders.
Therefore, the spectator mass $m_a$ can be set to zero from the beginning,
which simplifies the formulas considerably.

Before we consider the non-collinear-safe situation, we briefly
repeat the concept of the collinear-safe case described in
\citere{Dittmaier:2000mb}.
Following Eqs.~(4.24) and (4.27) from there,
the inclusive integral of $\gsub_{ia,\tau}(p_i,p_a,k)$ over $[\rd k]$ is 
proportional to 
\beq
\Gsub_{ia,\tau}(P_{ia}^2) = 
\int_{0}^{x_1} \rd x\, \cGsub_{ia,\tau}(P_{ia}^2,x) 
\eeq
with
\beq
\cGsub_{ia,\tau}(P_{ia}^2,x_{ia}) = 
-\frac{\bar P_{ia}^2}{2}
\int_{z_1(x_{ia})}^{z_2(x_{ia})} \rd z_{ia} \,
\gsub_{ia,\tau}(p_i,p_a,k),
\eeq
where we could set the lower limit $x_0$ of the $x_{ia}$-integration
to zero because of $m_a=0$.
Since, however, the squared lowest-order matrix element
$|\M_0|^2$ multiplying $\gsub_{ia,\tau}$ 
in Eq.~\refeq{eq:m2sub}
depends on the variable $x_{ia}$, the integration of 
$|\M_\sub|^2$ over $x=x_{ia}$ is performed employing a $[\dots]_+$
prescription,
\beqar
\lefteqn{
-\frac{\bar P_{ia}^2}{2} \int_{0}^{x_1} \rd x_{ia}\, 
\int_{z_1(x_{ia})}^{z_2(x_{ia})} \rd z_{ia} \, \gsub_{ia,\tau}(p_i,p_a,k)
\cdots } &\qquad&
\nn\\
&=& \int_{0}^{1} \rd x\, \left\{
\Gsub_{ia,\tau}(P_{ia}^2) \, \de(1-x) 
+ \left[\cGsub_{ia,\tau}(P_{ia}^2,x)\right]_+ \right\} \cdots.
\label{eq:Gia1}
\eeqar
This integration, where the ellipses stand for $x$-dependent functions
such as the squared lowest-order matrix elements and flux factors,
is usually done numerically. Since the soft and collinear singularities
occur at $x\to x_1=1-{\cal O}(m_\ga)$, the singular parts are entirely
contained in $\Gsub_{ia,\tau}(P_{ia}^2)$ in Eq.~\refeq{eq:Gia1},
and the upper limit $x_1$ could be replaced by $1$ in the actual
$x$-integration.
For completeness we give the explicit form of the functions
$\Gsub_{ia,\tau}$ and $\cGsub_{ia,\tau}$ in the limit $m_i\to0$,
\beq
\begin{array}[b]{rclrcl}
\Gsub_{ia,+}(P_{ia}^2) &=& 
\disp {\cal L}(|P_{ia}^2|,m_i^2) - \frac{\pi^2}{2} + 1,
&\qquad
\Gsub_{ia,-}(P_{ia}^2) &=& 
\disp \frac{1}{2},
\\[1em]
\cGsub_{ia,+}(P_{ia}^2,x)  &=&
\disp \frac{1}{1-x}\left[2\ln\biggl(\frac{2-x}{1-x}\biggr)-\frac{3}{2}\right],
&\qquad
\cGsub_{ia,-}(P_{ia}^2,x) &=& 0,
\end{array}
\label{eq:Giares}
\eeq
which are taken from Eqs.~(3.19) and (3.20) of
\citere{Dittmaier:2000mb}.

In a non-collinear-safe situation, the ellipses on the l.h.s.\ of
Eq.~\refeq{eq:Gia1} also involve $z_{ia}$-dependent functions,
as e.g.\ $\theta$-functions for cuts or event selection.
Thus, also the integration over $z_{ia}$ has to be performed 
numerically in this case, and we have to generalize Eq.~\refeq{eq:Gia1}
in an appropriate way.
To this end, we generalize the usual $[\dots]_+$ prescription in the
following way. Writing
\beq
\int \rd^n{\bf r}\, \Bigl[ g({\bf r}) \Bigr]^{(r_i)}_{+,(a)} \, f({\bf r}) 
\equiv
\int \rd^n{\bf r}\, g({\bf r}) 
\left( f({\bf r}) - f({\bf r})\Big|_{r_i=a} \right)
\eeq
for the $[\dots]_+$ prescription in the $r_i$-integration in
a multiple integral over $n$ variables $r_k$ $(k=1,\dots,n)$,
we can iterate this definition to two-dimensional integrals according to
\beqar
\int \rd^n{\bf r}\, \Bigl[ g({\bf r}) \Bigr]^{(r_i,r_j)}_{+,(a,b)} \, 
f({\bf r}) 
&\equiv&
\int \rd^n{\bf r}\, \biggl[ \Bigl[ g({\bf r}) \Bigr]^{(r_i)}_{+,(a)}
\biggr]^{(r_j)}_{+,(b)} \, f({\bf r}) 
\nn\\
&=&
\int \rd^n{\bf r}\, g({\bf r}) 
\left( f({\bf r}) - f({\bf r})\Big|_{r_i=a} - f({\bf r})\Big|_{r_j=b} 
+ f({\bf r})\Big|_{r_i=a \atop r_j=b} \right). \hspace{2em}
\eeqar
In the notation $[g({\bf r})]^{(r_i)}_{+,(a)}$ we omit the superscript
$(r_i)$ if $g({\bf r})$ depends only on the integration variable $r_i$,
and we omit the subscripts $(a)$ or $(a,b)$ if $a=1$ or $a=b=1$.
This obviously recovers the usual notation for the one-dimensional
prescription used above.
Introducing a double $[\dots]_+$ prescription in $x=x_{ia}$ and $z=z_{ia}$,
we generalize Eq.~\refeq{eq:Gia1} to
\beqar
\lefteqn{
-\frac{\bar P_{ia}^2}{2} \int_{0}^{x_1} \rd x\, 
\int_{z_1(x)}^{z_2(x)} \rd z \, \gsub_{ia,\tau}(p_i,p_a,k)
\cdots } &\qquad&
\nn\\*
&=& \int_{0}^{1} \rd x\, \int_{0}^{1} \rd z\, \left\{
\Gsub_{ia,\tau}(P_{ia}^2) \, \de(1-x) \, \de(1-z) 
+ \left[\cGsub_{ia,\tau}(P_{ia}^2,x)\right]_+ \de(1-z) \right.
\nn\\
&& \hspace{6em}\left. {}
+ \left[\bcGsub_{ia,\tau}(P_{ia}^2,z)\right]_+ \de(1-x)
+ \Bigl[\bar\gsub_{ia,\tau}(x,z)\Bigr]^{(x,z)}_+ 
\right\} \cdots.
\label{eq:Gia2}
\eeqar
If the functions hidden in the ellipses do not depend on $z$,
the last two terms within the curly brackets do not contribute
and the formula reduces to Eq.~\refeq{eq:Gia1}.

We derive Eq.~\refeq{eq:Gia2} and the explicit 
form of the two extra terms in two steps. 
In the derivation we quantify the
previous ellipses by the regular test function $f(x,z)$.
The first step introduces a $[\dots]_+$ prescription in the
$x$-integration of the l.h.s.\ of Eq.~\refeq{eq:Gia2} after
interchanging the order of the integrations,
\beqar
I[f] &\equiv&
-\frac{\bar P_{ia}^2}{2} \int_{0}^{x_1} \rd x\, 
\int_{z_1(x)}^{z_2(x)} \rd z \, \gsub_{ia,\tau} \, f(x,z)
\nn\\
&=& -\frac{\bar P_{ia}^2}{2}
\int_{0}^{1} \rd z\, \int_{0}^{x_1(z)} \rd x 
\, \gsub_{ia,\tau} \, f(x,z)
\nn\\
&=& -\frac{\bar P_{ia}^2}{2}
\int_{0}^{1} \rd z\, \int_{0}^{x_1(z)} \rd x \, 
\left\{ \Bigl[ \gsub_{ia,\tau} \Bigr]^{(x)}_{+,(x_1(z))} \, f(x,z)
+\gsub_{ia,\tau} \, \, f(x_1(z),z) \right\}. 
\label{eq:If1}
\eeqar
The upper limit $x_1(z)$ of the $x$-integration follows upon solving
the explicit form of the limits $z_{1,2}(x)$ (given in Eq.~(4.22)
of \citere{Dittmaier:2000mb}) for $x$. The full form of $x_1(z)$ is
rather complicated for finite $m_\ga$, but in the following it is
only needed for $m_\ga=0$, where it simplifies to
\beq
x_1(z)\Big|_{m_\ga=0} = \frac{\bar P_{ia}^2z}{\bar P_{ia}^2z-m_i^2(1-z)}.
\eeq
Note that soft or collinear singularities result from the region
of highest $x$ values, $x\to x_1=\max\{x_1(z)\}$, so that the
first term in curly brackets in Eq.~\refeq{eq:If1} is free of
such singularities owing to the $[\dots]_+$ regularization.
Thus, we can set $m_i\to0$ in this part, i.e.\ in particular $x_1(z)\to1$,
yielding
\beq
I[f] = -\frac{\bar P_{ia}^2}{2}
\int_{0}^{1} \rd z\, \left\{
\int_{0}^{1} \rd x \, \Bigl[ \gsub_{ia,\tau} \Bigr]^{(x)}_{+} \, f(x,z)
+ f(x_1(z),z) \int_{0}^{x_1(z)} \rd x \, \gsub_{ia,\tau}
\right\}.
\label{eq:If2}
\eeq
In the second step we introduce a $[\dots]_+$ prescription for the
$z$-integration in both terms,
\beqar
I[f] &=& -\frac{\bar P_{ia}^2}{2}
\int_{0}^{1} \rd z\, \Biggl\{
\int_{0}^{1} \rd x \,\biggl[ \Bigl[ \gsub_{ia,\tau} \Bigr]^{(x)}_{+} 
\biggr]^{(z)}_{+} \, f(x,z)
+ \int_{0}^{1} \rd x \, \Bigl[ \gsub_{ia,\tau} \Bigr]^{(x)}_{+} \, f(x,1)
\nn\\
&& \hspace{6em} {}
+ f(x_1(z),z) \, 
\left[ \int_{0}^{x_1(z)} \rd x \, \gsub_{ia,\tau} \right]^{(z)}_{+}
+ f(x_1(1),1) \int_{0}^{x_1(z)} \rd x \, \gsub_{ia,\tau}
\Biggr\}
\hspace{2em}
\nn\\[.5em]
&=& -\frac{\bar P_{ia}^2}{2}
\int_{0}^{1} \rd x \, \int_{0}^{1} \rd z\, 
\Bigl[ \gsub_{ia,\tau} \Bigr]^{(x,z)}_{+} \, f(x,z)
- \frac{\bar P_{ia}^2}{2}\int_{0}^{1} \rd x \, f(x,1) \,
\left[ \int_{0}^{1} \rd z\, \gsub_{ia,\tau} \right]^{(x)}_{+} 
\nn\\
&& {}
-\frac{\bar P_{ia}^2}{2}\int_{0}^{1} \rd z\, f(x_1(z),z) \, 
\left[ \int_{0}^{x_1(z)} \rd x \, \gsub_{ia,\tau} \right]^{(z)}_{+}
-\frac{\bar P_{ia}^2}{2}f(x_1(1),1) 
\int_{0}^{1} \rd z\, \int_{0}^{x_1(z)} \rd x \, \gsub_{ia,\tau}.
\nn\\
\label{eq:If3}
\eeqar
In the second equality we just reordered some factors and integrations.
Since all integrals over the test function $f$ are now free of singularities,
i.e.\ the singularities are contained in the integrals multiplying $f$,
we can set the regulator masses $m_\ga$ and $m_i$ to zero in the
arguments of $f$. Thus, we can write
\beqar
I[f] &=& 
\int_{0}^{1} \rd x \, \int_{0}^{1} \rd z\, 
\Bigl[ \bar\gsub_{ia,\tau}(x,z) \Bigr]^{(x,z)}_{+} \, f(x,z)
+\int_{0}^{1} \rd x \, f(x,1) \,
\left[ \cGsub_{ia,\tau}(P_{ia}^2,x) \right]_{+} 
\nn\\
&& {}
+\int_{0}^{1} \rd z\, f(1,z) \, 
\left[ \bcGsub_{ia,\tau}(P_{ia}^2,z) \right]_{+}
+f(1,1) \, \Gsub_{ia,\tau}(P_{ia}^2)
\label{eq:If4}
\eeqar
with the abbreviations
\beqar
\bar\gsub_{ia,\tau}(x,z) &=&
-\frac{\bar P_{ia}^2}{2} \, \gsub_{ia,\tau} \Big|_{m_\ga=0 \atop m_i=0},
\nn\\
\cGsub_{ia,\tau}(P_{ia}^2,x) &=&
- \frac{\bar P_{ia}^2}{2} \int_{0}^{1} \rd z\, \gsub_{ia,\tau}
\Big|_{m_\ga=0 \atop m_i=0},
\nn\\
\bcGsub_{ia,\tau}(P_{ia}^2,z) &=&
-\frac{\bar P_{ia}^2}{2} \int_{0}^{x_1(z)} \rd x \, \gsub_{ia,\tau} 
\Big|_{m_\ga=0},
\nn\\
\Gsub_{ia,\tau}(P_{ia}^2) &=&
-\frac{\bar P_{ia}^2}{2}
\int_{0}^{1} \rd z\, \int_{0}^{x_1(z)} \rd x \, \gsub_{ia,\tau}.
\label{eq:Giadefs}
\eeqar
Equation~\refeq{eq:If4} is equivalent to the anticipated result
\refeq{eq:Gia2}, which was to be shown.
The explicit results for $\cGsub_{ia,\tau}(P_{ia}^2,x)$ and 
$\Gsub_{ia,\tau}(P_{ia}^2)$ have already been given above
in Eq.~\refeq{eq:Giares}, the two remaining functions are easily 
evaluated to
\beqar
\bar \gsub_{ia,+}(x,z) &=&
\frac{1}{1-x}\left(\frac{2}{2-x-z}-1-z\right),
\qquad
\bar \gsub_{ia,-}(x,z) = 0,
\nn\\
\bcGsub_{ia,+}(P_{ia}^2,z) &=&
P_{ff}(z)\biggl[\ln\biggl(\frac{-P^2_{ia}z}{m_i^2}\biggr)-1\biggr]
-\frac{2\ln(2-z)}{1-z}+(1+z)\ln(1-z),
\nn\\
\bcGsub_{ia,-}(P_{ia}^2,z) &=& 1-z.
\eeqar
The collinear singularity $\propto\ln m_i$
that appears in non-collinear-safe observables
is contained in the function $\bcGsub_{ia,+}(P_{ia}^2,z)$.

The resulting $ia$ contribution
$|\M_{\sub,ia}(\Phi_1)|^2$ to the phase-space integral of the 
subtraction function reads
\beqar
\lefteqn{\int\rd\Phi_1\,|\M_{\sub,ia}(\Phi_1;\kappa_i)|^2
= -\frac{\alpha}{2\pi} Q_a\sigma_a Q_i\sigma_i 
\int_0^1\rd x\, 
\int\rd\tilde\Phi_{0,ia}(P_{ia}^2,x)\, \int_0^1 \rd z\, 
} &&
\nn\\*
&& {} \times 
\Theta_{\cut}\Bigl(p_i=z\tilde p_i(x),k=(1-z)\tilde p_i(x),\{\tilde k_n(x)\}\Bigr) \,
\nn\\*
&& {} \times 
\frac{1}{x} \,
\Biggl\{ 
\Gsub_{ia,\tau}(P_{ia}^2) \, \de(1-x) \, \de(1-z) 
+ \left[\cGsub_{ia,\tau}(P_{ia}^2,x)\right]_+ \de(1-z) 
\nn\\
&& \hspace{3em} {}
+ \left[\bcGsub_{ia,\tau}(P_{ia}^2,z)\right]_+ \de(1-x)
+ \Bigl[\bar \gsub_{ia,\tau}(x,z)\Bigr]^{(x,z)}_+ 
\Biggr\}
\left|\M_0\Big(\tilde p_i(x),\tilde p_a(x);\tau\kappa_i\Big)\right|^2,
\hspace{3em}
\label{eq:intia}
\eeqar
which generalizes Eq.~(3.18) of \citere{Dittmaier:2000mb}.
Again, the arguments of the step function $\Theta_{\cut}(p_i,k,\{\tilde k_n\})$
indicate on which momenta phase-space cuts are imposed.
We recall that $\tilde\Phi_{0,ia}$ is the phase space of momenta
$\tilde p_i(x)$ and $\{\tilde k_n(x)\}$ (without
final-state radiation) with rescaled incoming momentum $\tilde p_a(x)=xp_a$
instead of the original incoming momentum $p_a$.
In the actual evaluation of Eq.~\refeq{eq:intia}, thus, the two
phase-space points $\tilde\Phi_{0,ia}(P_{ia}^2,x)$ and
$\tilde\Phi_{0,ia}(P_{ia}^2,x=1)$ have to be generated for each value of $x$
owing to the plus prescription in $x$. The relevant value of the 
invariant $P_{ia}^2$ is then calculated separately via
$P_{ia}^2=(\tilde p_i-\tilde p_a)^2$ for each of the two points,
so that $P_{ia}^2$ results from the momenta entering the 
matrix element $\M_0$ in both cases.%
\footnote{For a more formal explanation of this subtle but important point we
refer to the discussion at the end of Section~6.3 of 
\citere{Catani:2002hc}.}
The variable $z$, however, is generated independently of the phase-space
points and does not influence the kinematics in the matrix element.

The combination of the subtraction procedures described in this and 
the previous section
has been successfully applied and compared to results obtained with
phase-space slicing in the calculations of electroweak corrections
to Drell--Yan-like W-boson production, $\Pp\Pp\to\PW\to\nu_l l+X$,
and to deep-inelastic neutrino scattering, 
$\nu_\mu N\to \nu_\mu/\mu+X$, building on the calculations
discussed in \citeres{Dittmaier:2001ay,Brensing:2007qm} 
and \cite{Diener:2003ss}, respectively.

\subsection{Phase-space slicing}

In the phase-space slicing approach the soft and collinear 
phase-space regions are excluded in the (numerical) integration 
of the squared amplitude of the real-emission process.
In the so-called two-cutoff slicing method the soft region is cut off
by demanding that the photon energy $k^0$ should be larger than a
lower cut $\De E$ which is much smaller than any relevant energy scale
of the process. The collinear regions are excluded by demanding that
each angle of the photon with any other direction of a light
charged particle should be larger than the cut value $\De\theta\ll1$.
Note that this phase-space splitting is not Lorentz invariant.
In the soft and collinear regions the photon phase space can be
integrated out analytically by employing the general factorization
properties of the squared amplitudes, which are, e.g., discussed
in Section~2.2 of \citere{Dittmaier:2000mb} (including polarization
effects). General results for the integral over the soft region 
can, e.g., be found in \citeres{'tHooft:1978xw,Denner:1993kt}.
The integrals over the collinear regions for final-state
radiation can be easily obtained from intermediate results of the two
previous sections as follows.

The cuts defining the collinear region for the photon--emitter system
of \refse{se:fsr-fin} translate into new limits for the integration
variables $y_{ij}$ and $z_{ij}$,
\beq
\frac{m_i^2(1-z_{ij})}{\bar P_{ij}^2 z_{ij}} < y_{ij} < 
\frac{(p_i^0)^2}{\bar P_{ij}^2} \frac{1-z_{ij}}{z_{ij}} \, \De\theta^2,
\qquad
0 < z_{ij} < 1-\frac{\De E}{p_i^0},
\eeq
which are asymptotically valid up to the relevant order in $m_i\to0$.
With these new limits on $y_{ij}$ we evaluate the integral defined in 
Eq.~\refeq{eq:bcGsubij} and obtain
\beq
\bcGsli_+(p_i^0,z) =
P_{ff}(z)\biggl[\ln\biggl(\frac{(p_i^0)^2\De\theta^2}{m_i^2}\biggr)-1\biggr],
\qquad
\bcGsli_-(p_i^0,z) = 1-z.
\eeq
The integrals of these functions over $z=z_{ij}$ are given by
\beq
\Gsli_+(p_i^0) =
-\biggl[\ln\biggl(\frac{\De E^2}{(p_i^0)^2}\biggr)+\frac{3}{2}\biggr]
\biggl[\ln\biggl(\frac{(p_i^0)^2\De\theta^2}{m_i^2}\biggr)-1\biggr],
\qquad
\Gsli_-(p_i^0) = \frac{1}{2}.
\eeq
As it should be, in these results the dependence on the spectator particle
$j$ completely disappears, because it was only needed in the phase-space
parametrization. We also note that the same results can be 
obtained from \refse{se:fsr-ini}, where the limits on $x_{ia}$ and
$z_{ia}$ are changed to
\beq
\frac{m_i^2(1-z_{ia})}{-\bar P_{ia}^2z_{ia}+m_i^2(1-z_{ia})}
< 1-x_{ia} <
\frac{(p_i^0)^2}{-\bar P_{ia}^2} \frac{1-z_{ia}}{z_{ia}} \, \De\theta^2,
\qquad
0 < z_{ia} < 1-\frac{\De E}{p_i^0}.
\eeq

Using the functions $\bcGsli_\tau$ and $\Gsli_\tau$, 
the integral over the collinear
photon emission cone around particle $i$ reads
\beqar
\int_{\mathrm{coll},i}\rd\Phi_1\,|\M_1(\Phi_1;\kappa_i)|^2
&=& \frac{\alpha}{2\pi} Q_i^2 \,
\int\rd\tilde\Phi_0\,
\int_0^1 \rd z\, 
\left\{ \Gsli_{\tau}(p_i^0) \de(1-z) 
+ \left[\bcGsli_{\tau}(p_i^0,z)\right]_+ \right\}
\nn\\[.3em]
&& {}\times
|\M_0(\tilde p_i;\tau\kappa_i)|^2 \,
\Theta_{\cut}\Bigl(p_i=z\tilde p_i,k=(1-z)\tilde p_i,\{k_n\}\Bigr),
\eeqar
where the momenta $\tilde p_i$ and $\{k_n\}$ belong to the
phase-space point $\tilde\Phi_0$.
Of course, apart from the polarization issue this is a well-known
result which can be found in various papers~\cite{Giele:1991vf}.%
\footnote{Descriptions of phase-space slicing for initial-state
radiation off unpolarized particles can also be found in
\citere{Giele:1991vf}; the case of polarized incoming particles
is, e.g., treated in \citere{Bohm:1993qx}.
}

\section{Collinear singularities from 
\boldmath{$\ga\to f\bar f^*$ splittings}}
\label{se:affst}

\subsection{Asymptotics in the collinear limit}

We consider a generic scattering process 
\beq
\gamma(k,\la_\ga)+a(p_a) \to f(p_f)+X,
\eeq
where the momenta of the particles are indicated in parentheses
and $\la_\ga=\pm$ is the photon helicity.
Here $a$ is any massless incoming particle and $f$ is an outgoing 
light fermion or antifermion. The remainder $X$ may contain
additional light fermions which can be treated in the same way as $f$.
For later use, we define the squared centre-of-mass energy $s$,
\beq
s = (p_a+k)^2 = 2p_a k.
\eeq
The collinear singularity in the squared matrix element
$|\M_{\gamma a\to fX}|^2$ occurs if the angle $\theta_f$ between $f$ and
the incoming $\gamma$ becomes small; in this limit the scalar product
$(kp_f)$ is of ${\cal O}(m_f^2)$, where $m_f$ is the small mass of $f$.
Neglecting terms that are irrelevant in the limit $m_f\to0$ the 
squared matrix element $|\M_{\gamma a\to fX}(k,p_a,p_f;\la_\ga)|^2$
for a definite photon helicity $\la_\ga=\pm$
(but summed over the polarizations of $f$)
asymptotically behaves like 
\beq
|\M_{\gamma a\to fX}(k,p_a,p_f;\la_\ga)|^2 \;\asymp{kp_f\to0}\;
Q_f^2e^2 \, h^{\ga f}_{\tau}(k,p_f) \,
|\M_{\bar fa\to X}(p_{\bar f}=xk,p_a;\kappa_{\bar f}=\tau\la_\ga)|^2,
\label{eq:affst-fact}
\eeq
where $x=1-p_f^0/k^0$ and $Q_fe$ is the electric charge of $f$.
The matrix element $\M_{\bar fa\to X}$ corresponds to the
related process $\bar fa\to X$ that results from 
$\gamma a(\to f\bar f^* a)\to fX$ upon cutting the $\bar f^*$ line
in all diagrams involving the splitting $\gamma\to f\bar f^*$
(see also \reffi{fig:affst}).
The incoming momenta relevant in the
different matrix elements are given in parentheses.
Moreover, in Eq.~\refeq{eq:affst-fact} we assume a summation over
$\tau=\pm$, where $\tau=\pm$ refers to the two cases where 
the sign $\kappa_{\bar f}$ of the $\bar f$ helicity
is equal or opposite to the photon helicity $\la_\ga$.
The functions $h^{\ga f}_{\tau}(k,p_f)$, which rule the structure of the
collinear singularity, are given by
\beqar
h^{\ga f}_{+}(k,p_f) &=& \frac{1}{x(kp_f)} \, 
\biggl(P_{f\ga}(x)
+\frac{xm_f^2}{kp_f}\biggr)
-h^{\ga f}_{-}(k,p_f),
\nn\\
h^{\ga f}_{-}(k,p_f) &=& \frac{1}{x(kp_f)} \,
(1-x)\biggl(1-x-\frac{m_f^2}{2kp_f}\biggr),
\label{eq:haf}
\eeqar
with the splitting function
\beq
P_{f\ga}(x) = (1-x)^2 +x^2.
\eeq
The derivation of this result is given in \refapp{app:affst-fac}.

Note that the collinear singularity for $kp_f\to0$ can be attributed
to a single external leg (namely $\bar f$) of the related hard process
$\bar fa\to X$. Thus, there is no need to construct the subtraction
function $|\M_{\sub}|^2$ from several dipole contributions 
$\propto Q_f Q_{f'}$. Instead we can construct $|\M_{\sub}|^2$ 
as a single term $\propto Q_f^2$. Nevertheless we select a spectator
$f'$ to the emitter $f$ for the phase-space construction, which
proceeds in complete analogy to the photon radiation case.
We have the freedom to choose any particle in the
initial or final state as spectator. In the following we
describe the ``dipole'' formalism in two variants: one with
a spectator from the initial state, another with a spectator
from the final state.
The two situations are illustrated in \reffi{fig:affst}.
\bfi
\centerline{
\begin{picture}(310,110)(0,0)
\put(0,-10){
  \begin{picture}(100,120)(0,0)
  \Photon(20,90)(50,70){2}{4}
  \Line(50,70)(80,50)
  \Line(80,50)( 40, 20)
  \LongArrow( 52,18)( 67, 30)
  \LongArrow( 65,70)( 80, 80)
  \LongArrow( 20,80)( 35, 70)
  \Line(50,70)(80,90)
  \Vertex( 50,70){2.5}
  \GCirc(80,50){10}{1}
  \put( 10, 88){$\gamma$}
  \put( 87, 90){$f$}
  \put( 52, 50){$\bar f$}
  \put( 28, 18){$a$}
  \put( 65, 18){$p_a$}
  \put( 20, 63){$k$}
  \put( 76, 68){$p_f$}
  \end{picture} } 
\put(130,-10){
  \begin{picture}(160,120)(0,0)
  \Photon(20,90)(50,70){2}{4}
  \Line(50,70)(80,50)
  \Line(80,50)(120, 20)
  \LongArrow( 90,30)(105, 18)
  \LongArrow( 65,70)( 80, 80)
  \LongArrow( 20,80)( 35, 70)
  \Line(50,70)(80,90)
  \Vertex( 50,70){2.5}
  \GCirc(80,50){10}{1}
  \put( 10, 88){$\gamma$}
  \put( 87, 90){$f$}
  \put( 52, 50){$\bar f$}
  \put(125, 15){$j$}
  \put( 90, 14){$p_j$}
  \put( 20, 63){$k$}
  \put( 76, 68){$p_f$}
  \end{picture} }
\end{picture} } 
\caption{Generic diagrams for the splittings $\ga\to f\bar f^*$ with 
an initial-state spectator $a$ or a final-state spectator $j$.}
\label{fig:affst}
\efi

\subsection{Initial-state spectator}
\label{se:affst-ini}

The function that is subtracted from the integrand
$|\M_{\gamma a\to fX}(k,p_a,p_f;\la_\ga)|^2$ is defined as follows,
\beq
|\M_{\sub}(\la_\ga)|^2 = Q_f^2 e^2 \, 
\hsub^{\ga f,a}_{\tau}(k,p_f,p_a) \,
\left|\M_{\bar f a\to X}\Bigl(\tilde p_{\bar f},p_a,
\{\tilde k_n\};\kappa_{\bar f}=\tau\la_\ga\Bigr)\right|^2,
\label{eq:Msubaffst-ini}
\eeq
with the radiator functions
\beqar
\hsub^{\ga f,a}_{+}(k,p_f,p_a) &=& \frac{1}{x_{f,\ga a}(kp_f)} \,
\biggl( P_{f\ga}(x_{f,\ga a})
+\frac{x_{f,\ga a}m_f^2}{kp_f}\biggr)
-\hsub^{\ga f,a}_{-}(k,p_f,p_a),
\nn\\
\hsub^{\ga f,a}_{-}(k,p_f,p_a) &=& \frac{1}{x_{f,\ga a}(kp_f)} \,
(1-x_{f,\ga a}) \biggl(1-x_{f,\ga a}-\frac{m_f^2}{2kp_f}\biggr),
\eeqar
and the auxiliary quantity
\beq
x_{f,\gamma a} = \frac{p_a k-p_f k-p_a p_f}{p_a k}.
\eeq
Here we kept the dependence on a finite $m_f$, because it is needed
in the analytical treatment of the singular phase-space integration below.
The modified momenta $\tilde p_{\bar f}$ and $\{\tilde k_n\}$
entering the squared matrix element on the r.h.s.\
of Eq.~\refeq{eq:Msubaffst-ini} will only be needed for $m_f=0$ in
applications with
small values of $m_f$. In this limit they can be chosen as
\beq
\tilde p_{\bar f}^\mu(x) = x k^\mu, \qquad
\tilde p_{\bar f}^\mu =
\tilde p_{\bar f}^\mu(x_{f,\gamma a}), \qquad
\tilde k_n^\mu = {\Lambda^\mu}_\nu \, k_n^\nu
\eeq
with the Lorentz transformation matrix ${\Lambda^\mu}_\nu$ given by
\beqar
{\Lambda^\mu}_\nu &=&
{g^\mu}_\nu - \frac{(P+\tilde P)^\mu(P+\tilde P)_\nu}{P^2+P\tilde P}
+\frac{2\tilde P^\mu P_{\nu}}{P^2}, 
\label{eq:Lambda}
\\
P^\mu &=& p_a^\mu+k^\mu-p_f^\mu, \qquad 
\tilde P^\mu(x)=p_a^\mu+\tilde p_{\bar f}^\mu(x), \qquad
\tilde P^\mu=p_a^\mu+\tilde p_{\bar f}^\mu.
\eeqar
It is straightforward to check that $|\M_{\sub}|^2$ possesses the same
asymptotic behaviour as $|\M_{\gamma a\to fX}|^2$
in Eq.~\refeq{eq:affst-fact} in the collinear limit with $m_f\to0$.
Thus, the difference $|\M_{\gamma a\to fX}|^2-|\M_{\sub}|^2$
can be integrated numerically for $m_f=0$.

The correct dependence of $|\M_{\sub}|^2$ (and the related kinematics)
on a finite $m_f$ is, however, needed when this function is integrated
over $\theta_f$ leading to the collinear singularity for $\theta_f\to0$.
The actual analytical integration can be done as described in
\citere{Dittmaier:2000mb} (even for finite $m_a$ and $m_f$).
Here we only sketch the individual steps and give the final result.
The $(N+1)$-particle phase space is first split into the corresponding
$N$-particle phase space and the integral over the remaining
degrees of freedom that contain the singularity,
\beq
\int\rd\phi(p_f,P;k+p_a) 
= \int_{0}^{x_1}\rd x
\int\rd\phi\Big(\tilde P(x);\tilde p_{\bar f}(x)+p_a\Big)
\int [\rd p_f(s,x,y_{f,\ga a})],
\eeq
with the explicit parametrization 
\beq
\int [\rd p_f(s,x,y_{f,\ga a})] = 
\frac{s}{4(2\pi)^3} \int_{y_1(x)}^{y_2(x)} \rd y_{f,\ga a} \,
\int \rd\phi_f.
\eeq
The upper kinematical limit of the parameter $x=x_{f,\ga a}$ is given by
\beq
x_1 = 1-\frac{2m_f}{\sqrt{s}},
\eeq
but in the limit $m_f\to0$ we can set $x_1=1$.
While the integration of the azimuthal angle $\phi_f$ of $f$ simply
yields a factor $2\pi$, the integration over the auxiliary parameter
\beq
y_{f,\ga a} = \frac{k p_f}{k p_a} = \frac{2k p_f}{s}
\eeq
with the boundary
\beq
y_{1,2}(x) = \frac{1}{2}\left[1-x \mp \sqrt{(1-x)^2-\frac{4m_f^2}{s}}\,\right]
\eeq
is less trivial.
Defining
\beq
\cHsub^{\ga f,a}_{\tau}(s,x) = \frac{xs}{2} 
\int_{y_1(x)}^{y_2(x)} \rd y_{f,\ga a} \, 
\hsub^{\ga f,a}_{\tau}(k,p_f,p_a),
\label{eq:cHsubgfa}
\eeq
the result of this straightforward integration (for $m_f\to0$) is
\beqar
\cHsub^{\ga f,a}_{+}(s,x) &=& P_{f \ga}(x)
  \ln\biggl(\frac{s(1-x)^2}{m_f^2}\biggr) + 2x(1-x)
-\cHsub^{\ga f,a}_{-}(s,x),
\nn\\
\cHsub^{\ga f,a}_{-}(s,x) &=& (1-x)^2
  \ln\biggl(\frac{s(1-x)^2}{m_f^2}\biggr) - (1-x)^2.
\eeqar
For clarity we finally give the
contribution $\sigma^{\sub}_{\gamma a\to fX}$ that has to be added
to the result for the cross section obtained from the integral of the
difference $|\M_{\gamma a\to fX}|^2-|\M_{\sub}|^2$,
\beq
\sigma^{\sub}_{\gamma a\to fX}(k,p_a;\la_\ga)
= N_{\mathrm{c},f}\,
\frac{Q_f^2\alpha}{2\pi} \int_0^1\rd x\, \cHsub^{\ga f,a}_{\tau}(s,x) \,
\sigma_{\bar fa\to X}(p_{\bar f}=xk,p_a;\kappa_{\bar f}=\tau\la_\ga).
\eeq
Although formulated for integrated cross sections, the previous
formula can be used to calculate any differential cross section
after obvious modifications.

For the case of unpolarized photons this subtraction variant has
already been briefly described in \citere{Diener:2003ss}, where it was
applied to the contributions to deep-inelastic neutrino scattering, 
$\nu_\mu N\to \nu_\mu/\mu+X$, that are induced by a photon
distribution function of the nucleon $N$. 
Moreover, the method presented here was successfully used in the calculation
of photon-induced real corrections to Drell--Yan-like W~production
(see Section~10 of \citere{SMH-LH2005} and \citere{Brensing:2007qm})
and of photon- and gluon-induced
real corrections to Higgs production via vector-boson fusion at the
LHC \cite{Ciccolini:2007jr}.
All these results were also cross-checked against phase-space slicing.

\subsection{Final-state spectator}
\label{se:affst-fin}

As an alternative to the case of an initial-state spectator
described in the previous section, we here present the treatment
with a possibly massive final-state spectator $j$ with mass $m_j$,
i.e.\ we consider the process
\beq
\gamma(k)+a(p_a) \to f(p_f)+j(p_j)+X.
\eeq
The initial-state particle $a$ is assumed massless in the following,
but all formulas can be generalized to $m_a\ne0$ following closely
the treatment of phase space described in Section~4.2 of
\citere{Dittmaier:2000mb}.
The subtraction function now is constructed as follows,
\beq
|\M_{\sub}(\la_\ga)|^2 = Q_f^2 e^2 \, 
\hsub^{\ga f}_{j,\tau}(k,p_f,p_j) \,
|\M_{\bar fa\to jX}(\tilde p_{\bar f},p_a,\tilde p_j;
\kappa_{\bar f}=\tau\la_\ga)|^2,
\label{eq:Msubaffst-fin}
\eeq
with the radiator functions
\beqar
\hsub^{\ga f}_{j,+}(k,p_f,p_j) &=& \frac{1}{x_{fj,\ga}(kp_f)} \,
\biggl(P_{f\ga}(x_{fj,\ga})
+\frac{x_{fj,\ga}m_f^2}{kp_f}\biggr)
-\hsub^{\ga f}_{j,-}(k,p_f,p_j),
\nn\\
\hsub^{\ga f}_{j,-}(k,p_f,p_j) &=& \frac{1}{x_{fj,\ga}(kp_f)} \,
(1-x_{fj,\ga})\biggl(1-x_{fj,\ga}-\frac{m_f^2}{2kp_f}\biggr)
\eeqar
and the auxiliary parameter
\beq
x_{fj,\ga} = \frac{k p_j+k p_f-p_f p_j}{k p_j + k p_f}.
\eeq
The momenta $\tilde p_{\bar f}$ and $\tilde p_j$ are given by
\beq
\tilde p_{\bar f}^\mu(x) = xk^\mu, \quad
\tilde p_{\bar f}^\mu=\tilde p_{\bar f}^\mu(x_{fj,\ga}), \quad
\tilde p_j^\mu = P^\mu + \tilde p_{\bar f}^\mu, \quad
P^\mu = p_f^\mu+p_j^\mu-k^\mu,
\label{eq:momaffst-fin}
\eeq
while the momenta of the other particles are unaffected.
Note that this construction of momenta is based on the restriction
$m_f=0$, which is used in the integration of the difference
$|\M_{\gamma a\to fjX}|^2-|\M_{\sub}|^2$.

In the integration of $|\M_{\sub}|^2$ over the collinear-singular
phase space, of course, the 
correct dependence on a finite $m_f$ is required. Owing to the
finite spectator mass $m_j$, this procedure is quite involved;
we sketch it in \refapp{app:affst-fin}.
Here we only present the results needed in practice.
The cross-section contribution $\sigma^{\sub}_{\gamma a\to fjX}$ 
that has to be added to the integrated
difference $|\M_{\gamma a\to fX}|^2-|\M_{\sub}|^2$ is given by
\beq
\sigma^{\sub}_{\gamma a\to fjX}(k,p_a;\la_\ga)
= N_{\mathrm{c},f}\,
\frac{Q_f^2\alpha}{2\pi} \int_0^1\rd x\, 
\cHsub^{\ga f}_{j,\tau}(P^2,x) \,
\sigma_{\bar fa\to jX}(p_{\bar f}=xk,p_a;\kappa_{\bar f}=\tau\la_\ga),
\label{eq:sigaffst-fin}
\eeq
where the collinear singularity is again contained in the kernels
\beqar
\cHsub^{\ga f}_{j,+}(P^2,x) &=& -P_{f\ga}(x)
  \ln\biggl[\frac{m_f^2 x}{(m_j^2-P^2)(1-x)}
    \biggl(1+\frac{m_j^2 x}{(m_j^2-P^2)(1-x)}\biggr)\biggr] 
     + 2x(1-x)
\nn\\
&& {} -\cHsub^{\ga f}_{j,-}(P^2,x),
\nn\\
\cHsub^{\ga f}_{j,-}(P^2,x) &=& -(1-x)^2
  \ln\biggl[\frac{m_f^2 x}{(m_j^2-P^2)(1-x)}
    \biggl(1+\frac{m_j^2 x}{(m_j^2-P^2)(1-x)}\biggr)\biggr] - (1-x)^2.
\nn\\
\label{eq:cHaffst-fin}
\eeqar
Of course, the singular contributions $\propto\ln m_f$ have the
same form as in the case of an initial-state spectator discussed in the
previous section.

\subsection{Phase-space slicing}

From the results of the two previous sections, the corresponding
formulas for the phase-space slicing approach can be easily obtained.
The collinear region, which is omitted in the phase-space integration,
is defined by the restriction $\theta_f<\De\theta$ on the
fermion emission angle $\theta_f$ in some given reference frame.

In \refse{se:affst-ini} this constraint translates into new limits
on the variable $y_{f,\ga a}$,
\beq
\frac{m_f^2}{s(1-x_{f,\gamma a})} < y_{f,\ga a} < 
\frac{(k^0)^2(1-x_{f,\gamma a})}{s}\De\theta^2,
\eeq
which modifies the result of the integral analogously defined to
Eq.~\refeq{eq:cHsubgfa} to
\beqar
\cHsli^{\ga f}_{+}(k^0,x) &=& P_{f \ga}(x)
  \ln\biggl(\frac{(k^0)^2(1-x)^2\De\theta^2}{m_f^2}\biggr) + 2x(1-x)
-\cHsub^{\ga f}_{-}(k^0,x),
\nn\\
\cHsli^{\ga f}_{-}(k^0,x) &=& (1-x)^2
  \ln\biggl(\frac{(k^0)^2(1-x)^2\De\theta^2}{m_f^2}\biggr) - (1-x)^2.
\eeqar
The cross-section contribution of the collinear region of $f$ then
reads
\beq
\sigma^{\mathrm{coll},f}_{\gamma a\to fX}(k,p_a;\la_\ga)
= N_{\mathrm{c},f}\,
\frac{Q_f^2\alpha}{2\pi} \int_0^1\rd x\, \cHsli^{\ga f}_{\tau}(k^0,x) \,
\sigma_{\bar fa\to X}(p_{\bar f}=xk,p_a;\kappa_{\bar f}=\tau\la_\ga).
\eeq

The same result is obtained from \refse{se:affst-fin} with
\refapp{app:affst-fin}, where the new limits on $z_{fj,\ga}$ read
\beq
\frac{m_f^2 x_{fj,\ga}}{-\bar P^2(1-x_{fj,\ga})} < 1-z_{fj,\ga} <
\frac{(k^0)^2x_{fj,\ga}(1-x_{fj,\ga})}{-\bar P^2}\De\theta^2.
\eeq

\section{Collinear singularities from 
\boldmath{$\ga^*\to f\bar f$ splittings}}
\label{se:astff}

\subsection{Asymptotics in the collinear limit}

We consider a generic scattering process 
\beq
a(p_a)+b(p_b) \to f(p_f)+\bar f(p_{\bar f})+X,
\eeq
where the momenta of the particles are indicated in parentheses.
Depending on the particle content of the remainder $X$, there may be
additional, independent collinear-singular configurations, but we 
are interested in the region where the invariant mass 
$(p_f+p_{\bar f})^2=2m_f^2+2p_fp_{\bar f}$
of the produced fermion--antifermion pair $f\bar f$ becomes of the
order ${\cal O}(m_f^2)$, where $m_f$ is small compared to typical
scales in the process.
The singular behaviour of the full squared matrix element
$|\M_{ab\to f\bar fX}(p_f,p_{\bar f})|^2$ entirely originates from 
diagrams containing a $\gamma^*\to f\bar f$ splitting, i.e.\
the singularity is related to the subprocess $ab\to\ga X$.
For the matrix element of this subprocess we write
$\M_{ab\to\ga X}=T_{ab\to\ga X}^\mu(\tilde k) \veps_{\la_\ga,\mu}(\tilde k)^*$,
where $T_{ab\to\ga X}^\mu(\tilde k)$ is the amplitude without the photon
polarization vector $\veps_{\la_\ga,\mu}(\tilde k)^*$. In the collinear limit
$p_f p_{\bar f}\to0$ the light-like
momentum $\tilde k$ is equal to $k=p_f+p_{\bar f}$
up to mass-suppressed terms. 
Neglecting terms that are irrelevant in the limit $m_f\to0$ the 
squared matrix element $|\M_{ab\to f\bar fX}(p_f,p_{\bar f})|^2$
asymptotically behaves like 
\beq
|\M_{ab\to f\bar fX}(p_f,p_{\bar f})|^2 \;\asymp{p_f p_{\bar f}\to0}\;
N_{\mathrm{c},f}\, Q_f^2e^2 \, h_{f\bar f,\mu\nu}(p_f,p_{\bar f}) \,
T_{ab\to\ga X}^\mu(\tilde k)^* \, T_{ab\to\ga X}^\nu(\tilde k),
\label{eq:astff-fact}
\eeq
where
\beq
h_{f\bar f,\mu\nu}(p_f,p_{\bar f}) = \frac{2}{(p_f+p_{\bar f})^2}
\left[ -g_{\mu\nu}+4z(1-z)\frac{k_{\perp,\mu}k_{\perp,\nu}}{k_\perp^2-m_f^2}
\right],
\qquad
z = \frac{p_f^0}{k^0},
\eeq
and $N_{\mathrm{c},f}$ is the colour multiplicity of $f$ 
($N_{\mathrm{c,lepton}}=1$, $N_{\mathrm{c,quark}}=3$).
The momentum $k_\perp$ is the component of $p_f$ that is orthogonal to the
collinear axis defined by $k$, i.e.\ $kk_\perp=0$,
and becomes of ${\cal O}(m_f)$ in the collinear limit.
An explicit prescription for the construction of $k_\perp$ can, e.g., be found
in \citere{Catani:2002hc}, where the analogous case of the gluonic
splitting into massive quarks $Q$, $\Pg^*\to Q\bar Q$, is worked out.
It is important to realize that $h_{f\bar f,\mu\nu}$ in 
Eq.~\refeq{eq:astff-fact}
is not proportional to the polarization sum 
$E_{\mu\nu}=\sum_{\la_\ga} \veps_{\la_\ga,\mu}(\tilde k)^* 
\veps_{\la_\ga,\nu}(\tilde k)$ of
the photon, so that the r.h.s.\ is not proportional to the
polarization-summed squared amplitude $|\M_{ab\to\ga X}|^2$
of the subprocess. This spin correlation has to be taken care of in the
construction of an appropriate subtraction function in order to
guarantee a point-wise cancellation of the singular behaviour in
the collinear phase-space region. The spin correlation encoded in
$h_{f\bar f,\mu\nu}$ drops out if the average over the azimuthal angle $\phi_f$
of the $\ga^*\to f\bar f$ splitting plane around the collinear axis
is taken.%
\footnote{As described in \citeres{Catani:1996jh,Catani:2002hc},
for unpolarized situations
this average can be easily obtained upon contraction with the projector
$\frac{1}{2}d_{\mu\nu}(k)=\frac{1}{2}[-g_{\mu\nu}+
(\mbox{``gauge terms'' involving $k^\mu$ or $k^\nu$})]/(1-\eps)$,
which fulfills $-g^{\mu\nu}d_{\mu\nu}(k)=2(1-\eps)$ and
$k^\mu d_{\mu\nu}(k)=0$ in $D=4-2\eps$ space-time dimensions.}
Indicating this averaging by 
$\langle\dots\rangle_{\phi_f}\equiv \int\rd\phi_f/(2\pi)$,
we get $\langle h_{f\bar f,\mu\nu}\rangle_{\phi_f} = E_{\mu\nu} h_{f\bar f}$
with (in four space-time dimensions)
\beq
h_{f\bar f}(p_f,p_{\bar f}) = \frac{2}{(p_f+p_{\bar f})^2}
\left[ P_{f\ga}(z)+\frac{2m_f^2}{(p_f+p_{\bar f})^2} \right]
\eeq
up to terms that are further suppressed by factors of $m_f$.
The averaged squared matrix element behaves as
\beq
\langle |\M_{ab\to f\bar fX}(p_f,p_{\bar f})|^2 \rangle_{\phi_f}
\;\asymp{p_f p_{\bar f}\to0}\;
N_{\mathrm{c},f}\,Q_f^2e^2 \, h_{f\bar f}(p_f,p_{\bar f}) \,
|\M_{ab\to\ga X}(\tilde k)|^2.
\label{eq:astff-fact2}
\eeq

Since the collinear singularity for $p_f p_{\bar f}\to0$ can be attributed
to a single external leg (the photon) of the related hard process
$ab\to\ga X$, also in this case
there is no need to construct the subtraction
function $|\M_{\sub}|^2$ from several dipole contributions. 
The function $|\M_{\sub}|^2$ can be chosen
as a single term $\propto Q_f^2$. Nevertheless a spectator
is selected for the phase-space construction, as in the previous section.
In the following we 
describe the ``dipole'' construction in two variants: one with
a spectator from the initial state, another with a spectator
from the final state.
The two situations are illustrated in \reffi{fig:astff}.
\bfi
\centerline{
\begin{picture}(310,110)(0,0)
\put(0,-10){
  \begin{picture}(100,120)(0,0)
  \Photon(20,50)(50,70){2}{4}
  \Line(50,70)(80, 90)
  \Line(20,50)(60, 20)
  \LongArrow(30,30)(45,18)
  \LongArrow(55,85)(70,95)
  \LongArrow(55,55)(70,45)
  \Line(50,70)(80,50)
  \Vertex(50,70){2.5}
  \GCirc(20,50){10}{1}
  \put(85,88){$f$}
  \put(85,46){$\bar f$}
  \put(30,70){$\gamma$}
  \put(65,15){$j$}
  \put(30,14){$p_j$}
  \put(55,99){$p_f$}
  \put(54,40){$p_{\bar f}$}
  \end{picture} } 
\put(130,-10){
  \begin{picture}(160,120)(0,0)
  \Photon(80,50)(110,70){2}{4}
  \Line(110,70)(140, 90)
  \Line(80,50)( 40, 20)
  \LongArrow( 52,18)( 67, 30)
  \LongArrow(115,85)(130, 95)
  \LongArrow(115,55)(130, 45)
  \Line(110,70)(140,50)
  \Vertex(110,70){2.5}
  \GCirc(80,50){10}{1}
  \put(145,88){$f$}
  \put(145,46){$\bar f$}
  \put(90,70){$\gamma$}
  \put( 28, 18){$a$}
  \put( 65, 18){$p_a$}
  \put(115,99){$p_f$}
  \put(114,40){$p_{\bar f}$}
  \end{picture} } 
\end{picture} } 
\caption{Generic diagrams for the splittings $\ga^*\to f\bar f$ with 
an initial-state spectator $a$ or a final-state spectator $j$,
where $f$ is a light fermion or antifermion.}
\label{fig:astff}
\efi

\subsection{Initial-state spectator}
\label{se:astff-ini}

We define the subtraction function as
\beq
|\M_{\sub}|^2 = N_{\mathrm{c},f}\,Q_f^2 e^2 \, 
h^a_{f\bar f,\mu\nu}(p_f,p_{\bar f},p_a) \,
T_{ab\to\ga X}^\mu(\tilde p_a,\tilde k)^* \, 
T_{ab\to\ga X}^\nu(\tilde p_a,\tilde k)
\label{eq:Msubastff-ini}
\eeq
with
\beq
h_{f\bar f}^{a,\mu\nu}(p_f,p_{\bar f},p_a) = 
\frac{2}{(p_f+p_{\bar f})^2} \left[-g^{\mu\nu}-\frac{4}{(p_f+p_{\bar f})^2}
\Bigl(z_{f\bar f,a} p^\mu_f-\bar z_{f\bar f,a}p^\mu_{\bar f}\Bigr)
\Bigl(z_{f\bar f,a} p^\nu_f-\bar z_{f\bar f,a}p^\nu_{\bar f}\Bigr) \right]
\eeq
and the auxiliary parameters
\beq
x_{f\bar f,a} = \frac{p_a p_f+p_a p_{\bar f}-p_f p_{\bar f}-m_f^2}%
	{p_a p_f+p_a p_{\bar f}},
\qquad
z_{f\bar f,a} = 1-\bar z_{f\bar f,a} = \frac{p_a p_f}{p_a p_f+p_a p_{\bar f}}. 
\label{eq:xzastff}
\eeq
The auxiliary momenta entering the amplitude for the related process
$ab\to\ga X$ are given by
\beqar
\tilde p_a^\mu(x) &=& x p_a^\mu, \qquad
\tilde p_a^\mu = \tilde p_a^\mu(x_{f\bar f,a}),
\nn\\
\tilde k^\mu(x) &=& P^\mu+\tilde p_a^\mu(x), \qquad 
\tilde k^\mu = \tilde k^\mu(x_{f\bar f,a}), \qquad 
P^\mu=p^\mu_f+p^\mu_{\bar f}-p_a^\mu,
\eeqar
while the momenta of the other particles remain unchanged.
In these equations we kept the dependence on $m_f$, but of course in
the numerical integration of 
$|\M_{ab\to f\bar fX}|^2-|\M_{\sub}|^2$ we can set $m_f$ to zero,
because we are only interested in the limit $m_f\to0$.
For the integration of $|\M_{\sub}|^2$ over the collinear-singular
region, we need the $m_f$-dependence of the spin average of
$h_{f\bar f}^{a,\mu\nu}$,
\beq
h^a_{f\bar f}(p_f,p_{\bar f},p_a) = 
\frac{2}{(p_f+p_{\bar f})^2} \left[ P_{f\ga}(z_{f\bar f,a})
+\frac{2m_f^2}{(p_f+p_{\bar f})^2} \right],
\eeq
and an appropriate phase-space splitting,
\beq
\int\rd\phi(p_f,p_{\bar f};P+p_a) = \int_{0}^{x_1}\rd x
\int\rd\phi\Big(\tilde k(x);P+\tilde p_a(x)\Big)
\int [\rd p_f(P^2,x,z)],
\eeq
where we have used the shorthands $x=x_{f\bar f,a}$ and $z=z_{f\bar f,a}$.
The explicit form of $\int[\rd p_f]$ reads
\beq
\int [\rd p_f(P^2,x,z)] = 
\frac{-P^2}{4(2\pi)^3} \,\frac{1}{x}\, \int_{z_1(x)}^{z_2(x)} \rd z \,
\int \rd\phi_f
\eeq
with the integration limits for the variables $x$ and $z$
\beq
x_1 = \frac{P^2}{P^2-4m_f^2},
\qquad
z_{1,2}(x) = \frac{1}{2}\left( 1\pm \sqrt{\frac{x_1-x}{x_1(1-x)}} \, \right).
\eeq
Separating the singular contributions as described in \refse{se:fsr-ini},
we rewrite the integral of $h^a_{f\bar f}$ for $m_f\to0$ as
\beqar
\lefteqn{
-\frac{P^2}{2} \int_{0}^{x_1} \rd x\, 
\int_{z_1(x)}^{z_2(x)} \rd z \, \hsub^a_{f\bar f}(p_f,p_{\bar f},p_a)
\cdots } &\qquad&
\nn\\
&=& \int_{0}^{1} \rd x\, \int_{0}^{1} \rd z\, \left\{
\Hsub^a_{f\bar f}(P^2) \, \de(1-x) \, \de(1-z) 
+ \left[\cHsub^a_{f\bar f}(P^2,x)\right]_+ \de(1-z) \right.
\nn\\
&& \hspace{6em}\left. {}
+ \left[\bcHsub^a_{f\bar f}(P^2,z)\right]_+ \de(1-x)
+ \Bigl[\bar h^a_{f\bar f}(x,z)\Bigr]^{(x,z)}_+ 
\right\} \cdots.
\label{eq:Hastff-ini}
\eeqar
The new functions $\bar h^a_{f\bar f}$, etc., defined here are obtained
from obvious substitutions and straightforward integrations,
\beqar
\bar h^a_{f\bar f}(x,z) &=& \frac{x}{1-x}P_{f\ga}(z),
\nn\\
\cHsub^a_{f\bar f}(P^2,x) &=& \frac{2x}{3(1-x)},
\nn\\
\bcHsub^a_{f\bar f}(P^2,z) &=& P_{f\ga}(z)
\biggl[\ln\biggl(\frac{-P^2 z(1-z)}{m_f^2}\biggr)-1\biggr] + 2z(1-z),
\nn\\
\Hsub^a_{f\bar f}(P^2) &=& \frac{2}{3} \ln\biggl(\frac{-P^2}{m_f^2}\biggr)
-\frac{16}{9}.
\eeqar
Using these functions the phase-space integral of the subtraction
function reads
\beqar
\lefteqn{\int\rd\Phi_{f\bar f}\,|\M_{\sub}(\Phi_{f\bar f})|^2
= N_{\mathrm{c},f}\, \frac{Q_f^2 \alpha}{2\pi}  \,
\int_0^1\rd x\, 
\int\rd\tilde\Phi_{\ga}(P^2,x)\, \int_0^1 \rd z\, 
} &&
\nn\\*
&& {} \times 
\Theta_{\cut}\Bigl(p_f=z\tilde k(x),p_{\bar f}=(1-z)\tilde k(x),
\{\tilde k_n(x)\}\Bigr) \,
\nn\\*
&& {} \times 
\frac{1}{x} \,
\Biggl\{ 
\Hsub^a_{f\bar f}(P^2) \, \de(1-x) \, \de(1-z) 
+ \left[\cHsub^a_{f\bar f}(P^2,x)\right]_+ \de(1-z) 
\nn\\
&& \hspace{3em} {}
+ \left[\bcHsub^a_{f\bar f}(P^2,z)\right]_+ \de(1-x)
+ \Bigl[\bar h^a_{f\bar f}(x,z)\Bigr]^{(x,z)}_+ 
\Biggr\}
\left|\M_{ab\to\ga X}\Big(\tilde p_a(x),\tilde k(x)\Big)\right|^2,
\hspace{3em}
\label{eq:intastff}
\eeqar
where we have made explicit which momenta enter the cut function 
$\Theta_{\cut}(p_f,p_{\bar f},\{\tilde k_n\})$.
Concerning the phase-space integration over $\rd\tilde\Phi_{\ga}(P^2,x)$
and its integration over the boost parameter $x$ the same comments
as made after Eq.~\refeq{eq:intia} apply. There are actually two
phase-space points for each $x$ value to be generated (one for $x<1$ and 
another for $x=1$), each determining momenta 
$\tilde p_a(x),\tilde k(x), \{\tilde k_n(x)\}$ for the evaluation
of $P^2$ and the matrix elements. The generation of the parameter $z$
proceeds independently, and
the squared amplitude $|\M_{ab\to\ga X}|^2$ in
Eq.~\refeq{eq:intastff} does not depend on $z$. 
Thus, if the full range in $z$ is integrated over, i.e.\ if the collinear
$f\bar f$ pair is treated as a single quasiparticle in the cut
procedure, the last two terms in curly brackets do not contribute.
In this case the fermion-mass logarithm is entirely contained in the
$\Hsub^a_{f\bar f}$ contribution. According to the KLN theorem
this contribution
will be completely compensated by virtual ${\cal O}(\alpha)$
corrections to the process $ab\to\ga X$ if collinear $f\bar f$ pairs
are not distinguished from emitted photons.

\subsection{Final-state spectator}
\label{se:astff-fin}

Since the case with a massive final-state spectator $j$ is quite
involved, we here present the formalism for $m_j=0$ and give
the details for the massive case in \refapp{app:astff-fin}.

\begin{sloppypar}
For $m_f=m_j=0$, the subtraction function can be defined as
\beq
|\M_{\sub}|^2 = N_{\mathrm{c},f}\,Q_f^2 e^2 \, 
h_{f\bar f,j,\mu\nu}(p_f,p_{\bar f},p_j) \,
T_{ab\to\ga jX}^\mu(\tilde k,\tilde p_j)^* \, 
T_{ab\to\ga jX}^\nu(\tilde k,\tilde p_j)
\label{eq:Msubastff-fin}
\eeq
with
\beq
h_{f\bar f,j}^{\mu\nu}(p_f,p_{\bar f},p_j) = 
\frac{2}{(p_f+p_{\bar f})^2} \left[-g^{\mu\nu}-\frac{2}{p_f p_{\bar f}}
\Bigl(z_{f\bar f j} p^\mu_f-\bar z_{f\bar fj}p^\mu_{\bar f}\Bigr)
\Bigl(z_{f\bar f j} p^\nu_f-\bar z_{f\bar fj}p^\nu_{\bar f}\Bigr) \right]
\eeq
and the auxiliary parameters
\beq
z_{f\bar f j} = 1-\bar z_{f\bar fj} = \frac{p_f p_j}{p_f p_j+p_{\bar f} p_j}, \qquad
y_{f\bar f j} = \frac{p_f p_{\bar f}}{p_f p_j+p_{\bar f} p_j+p_f p_{\bar f}}.
\label{eq:yzastff}
\eeq
The new momenta entering the amplitude for the related process
$ab\to\ga jX$ are given by
\beq
\tilde p_j^\mu = p_j^\mu/(1-y_{f\bar f j}), \qquad
\tilde k^\mu = P^\mu-\tilde p_j^\mu, \qquad 
P^\mu=p^\mu_f+p^\mu_{\bar f}+p_j^\mu,
\eeq
whereas all remaining momenta $k_n$ of particles in $X$ remain unchanged.
Equation~\refeq{eq:Msubastff-fin} can be used to integrate the difference
$|\M_{ab\to f\bar fjX}|^2-|\M_{\sub}|^2$ for massless fermions $f$.
In order to integrate $|\M_{\sub}|^2$ over the collinear-singular
region, the dependence on $m_f$ has to be taken into account. 
Details of this procedure can be found in \refapp{app:astff-fin}.
The result can be written in the form
\beqar
\int\rd\Phi_{f\bar f}\,|\M_{\sub}(\Phi_{f\bar f})|^2
&=& N_{\mathrm{c},f}\, \frac{Q_f^2\alpha}{2\pi} \,
\int\rd\tilde\Phi_{\ga}\,
\int_0^1 \rd z\, 
\Theta_{\cut}\Bigl(p_f=z\tilde k,
p_{\bar f}=(1-z)\tilde k,\tilde p_j,\{k_n\}\Bigr) 
\nn\\[.3em]
&& {}\times\left\{
\Hsub_{f\bar f,j}(P^2) \, \de(1-z) 
+ \left[\bcHsub_{f\bar f,j}(P^2,z)\right]_+ \right\}
|\M_{ab\to\gamma jX}(\tilde k,\tilde p_j)|^2
\nn\\
\label{eq:zintastff}
\eeqar
with
\beqar
\bcHsub_{f\bar f,j}(P^2,z) &=& P_{f\ga}(z) \left[ 
\ln\biggl(\frac{P^2z(1-z)}{m_f^2}\biggr)-1\right] + 2z(1-z),
\nn\\
\Hsub_{f\bar f,j}(P^2) &=&
\frac{2}{3}\ln\biggl(\frac{P^2}{m_f^2}\biggr) - \frac{16}{9}.
\label{eq:Hastff}
\eeqar
The momenta $\tilde k,\tilde p_j,\{k_n\}$ directly correspond to
the generated phase-space point in $\tilde\Phi_{\ga}$, while the
parameter $z$ is generated independently.
The comments on the $z$-integration made at the end of the previous 
subsection apply also here.
The squared amplitude $|\M_{ab\to\gamma jX}|^2$ in
Eq.~\refeq{eq:zintastff} does not depend on $z$, and
thus, if the event selection for $f$ and $\bar f$ is inclusive in 
the collinear region of the $\ga^*\to f\bar f$ splitting, 
the integral over $z$ trivially
reduces to the factor $\Hsub_{f\bar f,j}(P^2)$.
\end{sloppypar}

\subsection{Phase-space slicing}

Here we again deduce the integral over the collinear phase-space region
which is needed in the slicing approach. This region can, e.g., be defined
by restricting the angle $\theta_{f\bar f}$ between the $f$ and $\bar f$
directions to small values, $\theta_{f\bar f}<\De\theta\ll1$.

In \refse{se:astff-ini} this restriction leads to new limits in
$x_{f\bar f,a}$ and $z_{f\bar f,a}$,
\beq
\frac{m_f^2}{-P^2 z_{f\bar f,a}(1-z_{f\bar f,a})} < 1-x_{f\bar f,a} <
\frac{(k^0)^2}{-P^2} z_{f\bar f,a}(1-z_{f\bar f,a}) \De\theta^2,
\qquad
0 < z_{f\bar f,a} < 1,
\eeq
where $k^0=p_f^0+p_{\bar f}^0$ is the energy in the $f\bar f$ system.
This modifies the integrated results to
\beqar
\bcHsub_{f\bar f}(k^0,z) &=& P_{f\ga}(z)
\ln\biggl(\frac{(k^0)^2z^2(1-z)^2\De\theta^2}{m_f^2}\biggr) + 2z(1-z),
\nn\\
\Hsub_{f\bar f}(k^0) &=& 
\frac{2}{3} \ln\biggl(\frac{(k^0)^2\De\theta^2}{m_f^2}\biggr)
-\frac{23}{9},
\eeqar
where $\bcHsub_{f\bar f}$ and $\Hsub_{f\bar f}$ are defined 
analogously to Eq.~\refeq{eq:Hastff-ini}.
The integral of the squared matrix element over the collinear regions
then reads
\clearpage
\beqar
\int_{\theta_{f\bar f}<\De\theta}\rd\Phi_{f\bar f}\,
|\M_{ab\to {f\bar f}X}(\Phi_{f\bar f})|^2
&=& N_{\mathrm{c},f}\, \frac{Q_f^2\alpha}{2\pi} \,
\int\rd\tilde\Phi_{\ga}\,
\int_0^1 \rd z\, 
\nn\\
&& {}\times
\Theta_{\cut}\Bigl(p_f=z\tilde k,
p_{\bar f}=(1-z)\tilde k,\{\tilde k_n\}\Bigr) 
\\
&& {}\times
\left\{
\Hsub_{f\bar f}(k^0) \, \de(1-z) 
+ \left[\bcHsub_{f\bar f}(k^0,z)\right]_+ \right\}
|\M_{ab\to\gamma X}(\tilde k)|^2.
\nn
\eeqar

The same results can be obtained from \refse{se:astff-fin} with
\refapp{app:astff-fin},
where the new limits on the integration variables are given by
\beq
\frac{m_f^2}{\bar P^2} \, \frac{z_{f\bar fj}^2+(1-z_{f\bar fj})^2}%
{z_{f\bar fj}(1-z_{f\bar fj})} < y_{f\bar fj} <
\frac{(k^0)^2}{\bar P^2} z_{f\bar fj}(1-z_{f\bar fj}) \De\theta^2,
\qquad
0 < z_{f\bar fj} < 1.
\eeq

\section{Collinear singularities from 
\boldmath{$f\to f\ga^*$ splittings}}
\label{se:ffast}

\subsection{Asymptotics in the collinear limit}

We consider a generic scattering process 
\beq
f(p_f,\kappa_f)+a(p_a) \to f(p'_f)+X,
\eeq
with the momenta of the particles and the (sign of the)
helicity $\kappa_f=\pm$ of the incoming
fermion $f$ indicated in parentheses.
We are interested in the region where the squared momentum transfer
$(p_f-p'_f)^2=2m_f^2-2p_fp'_f$
of the scattered fermion $f$ becomes of the
order ${\cal O}(m_f^2)$, where $m_f$ is small compared to typical
scales in the process.
The singular behaviour of the full squared matrix element
$|\M_{fa\to fX}(p_f,p'_f;\kappa_f)|^2$ entirely originates from 
diagrams containing an $f\to f\gamma^*$ splitting, i.e.\
the singularity is related to the subprocess $\ga a\to X$.
For the matrix element of this subprocess we write
$\M_{\ga a\to X}(\tilde k,p_a,\la_\ga)=T_{\ga a\to X}^\mu(\tilde k) 
\veps_{\la_\ga,\mu}(\tilde k)$,
where $T_{\ga a\to X}^\mu(\tilde k)$ is the amplitude without the photon
polarization vector $\veps_{\la_\ga,\mu}(\tilde k)$. In the collinear limit
$p_f p'_f\to0$ the momentum $\tilde k$ is given by $k=p_f-p'_f$
up to mass-suppressed terms. 
Neglecting terms that are irrelevant in the limit $m_f\to0$ the 
squared matrix element $|\M_{fa\to fX}(p_f,p'_f;\kappa_f)|^2$
asymptotically behaves like 
\beq
|\M_{fa\to fX}(p_f,p_a,p'_f;\kappa_f)|^2 \;\asymp{p_f p'_f\to0}\;
N_{\mathrm{c},f}\, Q_f^2e^2 \, h^{ff}_{\kappa_f,\mu\nu}(p_f,p'_f) \,
T_{\ga a\to X}^\mu(\tilde k,p_a)^* \, T_{\ga a\to X}^\nu(\tilde k,p_a),
\label{eq:ffast-fact}
\eeq
where
\beqar
h^{ff}_{\kappa_f,\mu\nu}(p_f,p'_f) &=& \frac{-1}{(p_f-p'_f)^2} \Biggl[ -g_{\mu\nu}
-\frac{4(1-x)}{x^2}\frac{k_{\perp,\mu}k_{\perp,\nu}}{k_\perp^2-x^2 m_f^2}
\label{eq:hff}
\\
&& \hspace{3em} {}
+ \frac{\kappa_f}{x}\Biggl(2-x+\frac{2x^2 m_f^2}{(p_f-p'_f)^2}\Biggr)
\Bigl( \veps_{+,\mu}(\tilde k)^*\veps_{+,\nu}(\tilde k)
-\veps_{-,\mu}(\tilde k)^*\veps_{-,\nu}(\tilde k) \Bigr)
\Biggr]
\nn
\eeqar
with
\beq
x = \frac{k^0}{p_f^0}.
\eeq
The momentum $k_\perp$ is the component of $k$ that is orthogonal to the
collinear axis defined by $p_f$, i.e.\ $k_\perp p_f=0$,
and becomes of ${\cal O}(m_f)$ in the collinear limit.
A derivation of this factorization is described in \refapp{app:ffast-fac}.
Note that $h^{ff}_{\kappa_f,\mu\nu}$ in Eq.~\refeq{eq:ffast-fact}
is not proportional to the polarization sum 
$E_{\mu\nu}=\sum_{\la_\ga} \veps_{\la_\ga,\mu}(\tilde k)^* 
\veps_{\la_\ga,\nu}(\tilde k)$ of
the photon, so that the r.h.s.\ is not proportional to the
polarization-summed squared amplitude $|\M_{\ga a\to X}|^2$
of the subprocess. This spin correlation has to be taken into account in the
construction of an appropriate subtraction function in order to
guarantee a point-wise cancellation of the singular behaviour in
the collinear phase-space region. The spin correlation encoded in
$h^{ff}_{\kappa_f,\mu\nu}$ 
drops out if the average over the azimuthal angle $\phi'_f$
of the $f\to f\ga$ splitting plane around the collinear axis
is taken.
Details of this averaging process, which is indicated by
$\langle\dots\rangle_{\phi'_f}$, are given in \refapp{app:ffast-fac}.
The result is
\beq
\langle |\M_{fa\to fX}(p_f,p_a,p'_f;\kappa_f)|^2 \rangle_{\phi'_f}
\;\asymp{p_f p'_f\to0}\;
N_{\mathrm{c},f}\,Q_f^2e^2 \, h^{ff}_{\tau}(p_f,p'_f) \,
|\M_{\ga a\to X}(\tilde k,p_a;\la_\ga=\tau\kappa_f)|^2
\label{eq:ffast-fact2}
\eeq
with summation over $\tau=\pm$ and
\beq
h^{ff}_{\tau}(p_f,p'_f) = \frac{-1}{x(p_f-p'_f)^2}
\left[ P_{\ga f}(x) +\frac{2x m_f^2}{(p_f-p'_f)^2} 
+\tau\left(2-x+\frac{2x^2 m_f^2}{(p_f-p'_f)^2}\right)
\right],
\label{eq:hff2}
\eeq
which is valid in four space--time dimensions
up to terms that are further suppressed by factors of $m_f$.
Here $P_{\ga f}(x)$ is the splitting function
\beq
P_{\ga f}(x) = \frac{1+(1-x)^2}{x}.
\eeq

Since the collinear singularity for $p_f p'_f\to0$ can be attributed
to a single leg (the photon) of the related hard process
$\ga a\to X$, also in this case
there is no need to construct the subtraction
function $|\M_{\sub}|^2$ from several dipole contributions. 
The function $|\M_{\sub}|^2$ can be chosen
as a single term $\propto Q_f^2$, and a spectator
is only used in the phase-space construction as previously.
In the following we again
describe the ``dipole'' construction in two variants: one with
a spectator from the initial state, another with
a spectator from the final state.
The two situations are illustrated in \reffi{fig:ffast}.
\bfi
\centerline{
\begin{picture}(310,110)(0,0)
\put(0,-10){
  \begin{picture}(100,120)(0,0)
  \Line(20,90)(50,70)
  \Photon(50,70)(80,50){2}{4}
  \Line(80,50)( 40, 20)
  \LongArrow( 52,18)( 67, 30)
  \LongArrow( 65,70)( 80, 80)
  \LongArrow( 20,80)( 35, 70)
  \Line(50,70)(80,90)
  \Vertex( 50,70){2.5}
  \GCirc(80,50){10}{1}
  \put(  8, 88){$f$}
  \put( 87, 90){$f$}
  \put( 52, 50){$\gamma$}
  \put( 28, 18){$a$}
  \put( 65, 18){$p_a$}
  \put( 20, 63){$p_f$}
  \put( 78, 68){$p'_f$}
  \end{picture} } 
\put(130,-10){
  \begin{picture}(160,120)(0,0)
  \Line(20,90)(50,70)
  \Photon(50,70)(80,50){2}{4}
  \Line(80,50)(120, 20)
  \LongArrow( 90,30)(105, 18)
  \LongArrow( 65,70)( 80, 80)
  \LongArrow( 20,80)( 35, 70)
  \Line(50,70)(80,90)
  \Vertex( 50,70){2.5}
  \GCirc(80,50){10}{1}
  \put(  8, 88){$f$}
  \put( 87, 90){$f$}
  \put( 52, 50){$\gamma$}
  \put(125, 15){$j$}
  \put( 90, 14){$p_j$}
  \put( 20, 63){$p_f$}
  \put( 78, 68){$p'_f$}
  \end{picture} }
\end{picture} } 
\caption{Generic diagrams for the splittings $f\to f\ga^*$ with 
an initial-state spectator $a$ or a final-state spectator $j$,
where $f$ is a light fermion or antifermion.}
\label{fig:ffast}
\efi

\subsection{Initial-state spectator}
\label{se:ffast-ini}

We define the subtraction function as
\beq
|\M_{\sub}(\kappa_f)|^2 = N_{\mathrm{c},f}\,Q_f^2 e^2 \, 
h^{ff,a}_{\kappa_f,\mu\nu}(p_f,p'_f,p_a) \,
T_{\ga a\to X}^\mu(\tilde k,p_a)^* \, 
T_{\ga a\to X}^\nu(\tilde k,p_a)
\label{eq:Msubffast-ini}
\eeq
with
\beqar
h_{\kappa_f}^{ff,a,\mu\nu}(p_f,p'_f,p_a) &=& 
\frac{-1}{(p_f-p'_f)^2} \Biggl[-g^{\mu\nu}-\frac{4(1-x_{f,fa})}{x_{f,fa}^2}
\frac{\tilde k_\perp^\mu\tilde k_\perp^\nu}{\tilde k_\perp^2-x_{f,fa}^2 m_f^2}
\\
&& \hspace{3em} {}
+ \frac{\kappa_f}{x}\Biggl(2-x+\frac{2x^2 m_f^2}{(p_f-p'_f)^2}\Biggr)
\Bigl( \veps^\mu_+(\tilde k)^*\veps^{\nu}_+(\tilde k)
-\veps^\mu_-(\tilde k)^*\veps^{\nu}_-(\tilde k) \Bigr)
\Biggr]
\nn
\eeqar
and the auxiliary parameters
\beq
x_{f,fa} = \frac{p_a p_f-p_f p'_f-p_a p'_f+m_f^2}{p_a p_f},
\qquad
y_{f,fa} = \frac{p_f p'_f-m_f^2}{p_a p_f}. 
\label{eq:xyffast}
\eeq
Assuming again the incoming particle $a$ to be massless and
defining
\beq
s = (p_f+p_a)^2 = m_f^2+2p_a p_f = \bar s+m_f^2, \qquad
P^\mu = p_f^\mu+p_a^\mu-p_f^{\prime \mu},
\eeq
the needed auxiliary momenta for the related process
$\ga a\to X$ are given by
\beqar
\tilde k^\mu(x) &=& x\biggl(p_f^\mu-\frac{m_f^2}{\bar s}p_a^\mu\biggr), \qquad 
\tilde k^\mu = \tilde k^\mu(x_{f,fa}),\qquad
\tilde k_\perp^\mu = p_f^{\prime \mu} 
-\frac{p'_f \tilde k}{p_a \tilde k}\, p_a^\mu,
\nn\\
\tilde P^\mu(x)&=&\tilde k^\mu(x)+p_a^\mu, \qquad
\tilde P^\mu = \tilde P^\mu(x_{f,fa}), 
\nn\\[.3em]
\tilde k_n^\mu &=& {\Lambda^\mu}_\nu \, k_n^\nu,
\eeqar
where the Lorentz transformation matrix ${\Lambda^\mu}_\nu$ is
constructed from the momenta $P^\mu$ and $\tilde P^\mu$ as in
Eq.~\refeq{eq:Lambda}.
In these equations we kept the dependence on $m_f$, but of course in
the numerical integration of 
$|\M_{fa\to fX}|^2-|\M_{\sub}|^2$ we can set $m_f$ to zero if we
are only interested in the limit $m_f\to0$.
For the integration of $|\M_{\sub}|^2$ over the collinear-singular
region, we need the $m_f$-dependence of its azimuthal average,
\beq
\langle |\M_{\sub}(\kappa_f)|^2 \rangle_{\phi'_f} =
N_{\mathrm{c},f}\,Q_f^2e^2 \, 
h^{ff,a}_{\tau}(p_f,p'_f,p_a) \,
|\M_{\ga a\to X}(\tilde k,p_a;\la_\ga=\tau\kappa_f)|^2
\eeq
with summation over $\tau=\pm$ and
\beq
h^{ff,a}_{\tau}(p_f,p'_f,p_a) = 
\frac{1}{\bar s xy} \left[ P_{\ga f}(x) - \frac{2x(1-x)m_f^2}%
{y[\bar s(1-x-y)-m_f^2(2x+y)]}
+\tau\left(2-x-\frac{2x^2 m_f^2}{\bar sy}\right)
\right],
\eeq
where we have used the shorthands $x=x_{f,fa}$ and $y=y_{f,fa}$.
An appropriate phase-space splitting is given by
\beq
\int\rd\phi(p'_f,P;p_f+p_a) = \int_{0}^{x_1}\rd x
\int\rd\phi\Big(\tilde P(x);\tilde k(x)+p_a\Big)
\int [\rd p'_f(s,x,y)]
\eeq
with the explicit form of $\int[\rd p'_f]$ 
\beq
\int [\rd p'_f(s,x,y)] = 
\frac{\bar s}{4(2\pi)^3} \int_{y_1(x)}^{y_2(x)} \rd y \,
\int \rd\phi'_f
\eeq
and the integration limits for the variables $x$ and $y$
\beq
x_1 = \frac{\sqrt{s}-m_f}{\sqrt{s}+m_f},
\qquad
y_{1,2}(x) = \frac{\bar s}{2s}\left( 1-x-\frac{2m_f^2}{\bar s}x
\mp \sqrt{(1-x)^2-\frac{4m_f^2}{\bar s}x} \, \right).
\eeq
In the limit $m_f\to0$ the integral
\beq
\cHsub^{ff,a}_\tau(s,x) = \frac{x\bar s}{2} \int_{y_1(x)}^{y_2(x)} \rd y \,
h^{ff,a}_{\tau}(p_f,p'_f,p_a) 
\label{eq:Hffast-ini}
\eeq
can be easily evaluated to
\beq
\cHsub^{ff,a}_\tau(s,x) = \ln\biggl(\frac{\sqrt{s}(1-x)}{x m_f}\biggr)
\Bigl[ P_{\ga f}(x) + \tau (2-x)\Bigr] - \frac{1-x}{x} - \tau(1-x),
\eeq
and the part to be added to the cross section reads
\beq
\sigma^{\sub}_{fa\to fX}(p_f,p_a;\kappa_f) = 
\frac{Q_f^2\alpha}{2\pi} \int_0^1\rd x\, 
\cHsub^{ff,a}_{\tau}(s,x) \,
\sigma_{\gamma a\to X}(\tilde k=xp_f,p_a;\la_\ga=\tau\kappa_f).
\label{eq:sigffast-ini}
\eeq

\subsection{Final-state spectator}
\label{se:ffast-fin}

As an alternative to the case of an initial-state spectator,
we now present the treatment with a final-state spectator $j$,
i.e.\ we consider the process
\beq
f(p_f,\kappa_f)+a(p_a) \to f(p'_f)+j(p_j)+X.
\eeq
The particles $a$ and $j$ are assumed massless in the following;
the case of a massive spectator $j$ is described in \refapp{app:ffast-fin}.
The subtraction function is constructed as follows,
\beq
|\M_{\sub}(\kappa_f)|^2 = N_{\mathrm{c},f}\,Q_f^2 e^2 \, 
h^{ff}_{j,\kappa_f,\mu\nu}(p_f,p'_f,p_j) \,
T_{\ga a\to jX}^\mu(\tilde k,p_a,\tilde p_j)^* \, 
T_{\ga a\to jX}^\nu(\tilde k,p_a,\tilde p_j)
\label{eq:Msubffast-fin}
\eeq
with
\beqar
h_{j,\kappa_f}^{ff,\mu\nu}(p_f,p'_f,p_j) &=& 
\frac{-1}{(p_f-p'_f)^2} \Biggl[-g^{\mu\nu}
-\frac{4(\bar z_{fj,f} p^{\prime \mu}_f-z_{fj,f}p^\mu_j)
(\bar z_{fj,f} p^{\prime \nu}_f-z_{fj,f}p^\nu_j) 
}{(p_f-p'_f)^2 x_{fj,f}^2 \bar z_{fj,f}}
\\
&& \hspace{3em} {}
+ \frac{\kappa_f}{x}\Biggl(2-x+\frac{2x^2 m_f^2}{(p_f-p'_f)^2}\Biggr)
\Bigl( \veps^\mu_+(\tilde k)^*\veps^{\nu}_+(\tilde k)
-\veps^\mu_-(\tilde k)^*\veps^{\nu}_-(\tilde k) \Bigr)
\Biggr]
\nn
\eeqar
and the auxiliary parameters
\beq
x_{fj,f} = \frac{p_f p'_f+p_f p_j-p'_f p_j}{p_f p'_f+p_f p_j},
\qquad
z_{fj,f} = 1-\bar z_{fj,f} = \frac{p_f p'_f}{p_f p'_f+p_f p_j}.
\label{eq:xzffast}
\eeq
The momenta $\tilde k$ and $\tilde p_j$ are given by
\beq
\tilde k^\mu=x_{fj,f}p_f^\mu, \quad
\tilde p_j = P^\mu + \tilde k^\mu, \quad
P^\mu = p_f^{\prime \mu}+p_j^\mu-p_f^\mu.
\label{eq:momffast-fin}
\eeq
Note that this construction of momenta is based on the resctriction
$m_f=0$, which is used in the integration of the difference
$|\M_{fa\to fjX}|^2-|\M_{\sub}|^2$ for $m_f\to0$.

In the integration of $|\M_{\sub}|^2$ over the collinear-singular
phase space the correct dependence on a finite $m_f$ is required. 
We sketch this procedure in \refapp{app:ffast-fin} for a possibly
finite spectator mass $m_j$, but here we give only the relevant
formulas needed in applications.
The cross-section contribution $\sigma^{\sub}_{fa\to fjX}$ 
that has to be added to the integrated
difference $|\M_{fa\to fX}|^2-|\M_{\sub}|^2$ is given by
\beq
\sigma^{\sub}_{fa\to fX}(p_f,p_a;\kappa_f) = 
\frac{Q_f^2\alpha}{2\pi} \int_0^1\rd x\, 
\cHsub_{ffj,\tau}(P^2,x) \,
\sigma_{\gamma a\to jX}(\tilde k=xp_f,p_a;\la_\ga=\tau\kappa_f),
\label{eq:sigffast-fin}
\eeq
where the collinear singularity is contained in the kernels
\beq
\cHsub_{ffj,\tau}(P^2,x) = \frac{1}{2}
\ln\biggl(\frac{-P^2(1-x)}{x^3 m_f^2}\biggr)
\Bigl[ P_{\ga f}(x) + \tau (2-x)\Bigr] - \frac{1-x}{x} - \tau(1-x).
\eeq

\subsection{Phase-space slicing}

Finally, we derive the integral over the collinear phase-space region
for the slicing approach. This region is defined
by restricting the angle $\theta'_f$ between the outgoing and incoming
$f$ to small values, $\theta'_f<\De\theta\ll1$.

In \refse{se:ffast-ini} this restriction leads to new limits in
$y_{f,fa}$,
\beq
\frac{m_f^2}{s}\frac{x_{f,fa}^2}{1-x_{f,fa}} < y_{f,fa} <
\frac{(p_f^0)^2}{s} (1-x_{f,fa}) \De\theta^2,
\eeq
which modify the integrated result to
\beq
\cHsli^{ff}_\tau(p_f^0,x) = \ln\biggl(\frac{p_f^0(1-x)\De\theta}{x m_f}\biggr)
\Bigl[ P_{\ga f}(x) + \tau (2-x)\Bigr] - \frac{1-x}{x} - \tau(1-x),
\eeq
where the integral is defined analogously to Eq.~\refeq{eq:Hffast-ini}.
The cross-section contribution for the collinear scattering of $f$
is given by
\beq
\sigma^{\mathrm{coll},f}_{fa\to fX}(p_f,p_a;\kappa_f) = 
\frac{Q_f^2\alpha}{2\pi} \int_0^1\rd x\, 
\cHsli^{ff}_{\tau}(p_f^0,x) \,
\sigma_{\gamma a\to X}(\tilde k=xp_f,p_a;\la_\ga=\tau\kappa_f).
\eeq
The same results can be obtained from \refse{se:ffast-fin} with
\refapp{app:ffast-fin},
where the new limits on the integration variables are given by
\beq
\frac{m_f^2}{-\bar P^2} 
\frac{x_{fj,f}[1+(1-x_{fj,f})^2]}{1-x_{fj,f}}
< z_{fj,f} <
\frac{(p_f^0)^2}{-\bar P^2} x_{fj,f}(1-x_{fj,f}) \De\theta^2.
\eeq

\section{\boldmath{Application to the process
$\Pem\ga\to\Pem\mu^-\mu^+$}}
\label{se:appl}

In this section we illustrate the application of the methods
described in \refses{se:affst}, \ref{se:astff}, and
\ref{se:ffast} to the process $\Pem\ga\to\Pem\mu^-\mu^+$ at a
centre-of-mass energy $\sqrt{s}$ much larger than the involved
particle masses, $\sqrt{s}\gg\Me,m_\mu$. Of course, this process is
not of particular importance in particle phenomenology, but it
involves the three issues of (i) incoming photons splitting
into light $f\bar f$ pairs, (ii) the collinear production of
light $f\bar f$ pairs, and (iii) forward-scattered fermions 
and, thus, provides a good test process for these
cases. As already mentioned in \refse{se:fsr}, our treatment of
non-collinear-safe final-state radiation has already been tested
in other processes.

To illustrate the formalism, it is sufficient to consider the process
$\Pem\ga\to\Pem\mu^-\mu^+$ in QED, where only the four diagrams
shown in \reffi{fig:diags} contribute.
\bfi
\centerline{
{\unitlength .5pt \scriptsize
\begin{picture}(130,120)(0,0)
\SetScale{.5}
\ArrowLine(15, 95)( 65,90)
\Photon( 65, 10)(15, 5){4}{4}
\ArrowLine( 65, 10)( 75,50)
\Photon( 65,90)( 75, 50){4}{3.5}
\ArrowLine( 75, 50)(125,50)
\ArrowLine(125,  5)( 65, 10)
\ArrowLine( 65,90)(125,95)
\Vertex(65,90){3}
\Vertex(65,10){3}
\Vertex(75,50){3}
\put(-5,90){$\Pe^-$}
\put(-5, 0){$\gamma$}
\put(50,65){$\gamma$}
\put(50,25){$\mu$}
\put(130, 90){{$\Pe^-$}}
\put(130,45){{$\mu^-$}}
\put(130, 0){{$\mu^+$}}
\SetScale{1}
\end{picture}
}
\hspace*{2em}
{\unitlength .5pt \scriptsize
\begin{picture}(130,120)(0,0)
\SetScale{.5}
\ArrowLine(15, 95)( 65,90)
\Photon( 65, 10)(15, 5){4}{4}
\ArrowLine( 75,50)( 65, 10)
\Photon( 65,90)( 75, 50){4}{3.5}
\ArrowLine(125,50)( 75, 50)
\ArrowLine( 65, 10)(125,  5)
\ArrowLine( 65,90)(125,95)
\Vertex(65,90){3}
\Vertex(65,10){3}
\Vertex(75,50){3}
\put(-5,90){$\Pe^-$}
\put(-5, 0){$\gamma$}
\put(50,65){$\gamma$}
\put(50,25){$\mu$}
\put(130, 90){{$\Pe^-$}}
\put(130,45){{$\mu^+$}}
\put(130, 0){{$\mu^-$}}
\SetScale{1}
\end{picture}
}
\hspace*{2em}
{\unitlength .5pt \scriptsize
\begin{picture}(130,120)(0,0)
\SetScale{.5}
\ArrowLine(15,95)(55,70)
\Photon(55,10)(15, 5){4}{3.5}
\ArrowLine(55,70)(55,10)
\Photon(95,70)( 55,70){4}{3}
\ArrowLine(125,90)(95,70)
\ArrowLine(95,70)(125,45)
\ArrowLine( 55,10)(125, 5)
\Vertex(55,70){3}
\Vertex(55,10){3}
\Vertex(95,70){3}
\put(-5,90){$\Pe^-$}
\put(-5, 0){$\gamma$}
\put(38,37){$\Pe$}
\put(65,85){$\gamma$}
\put(130,88){{$\mu^+$}}
\put(130,43){{$\mu^-$}}
\put(130, 0){{$\Pe^-$}}
\SetScale{1}
\end{picture}
}
\hspace*{2em}
{\unitlength .5pt \scriptsize
\begin{picture}(160,120)(0,0)
\SetScale{.5}
\ArrowLine(15,95)(55,50)
\Photon(55,50)(15, 5){4}{4.5}
\ArrowLine(55,50)(95,50)
\Photon(95,50)(125,70){4}{3}
\ArrowLine(155,90)(125,70)
\ArrowLine(125,70)(155,45)
\ArrowLine( 95,50)(155, 5)
\Vertex(55,50){3}
\Vertex(95,50){3}
\Vertex(125,70){3}
\put(-5,90){$\Pe^-$}
\put(-5, 0){$\gamma$}
\put(73,35){$\Pe$}
\put(98,70){$\gamma$}
\put(160,88){{$\mu^+$}}
\put(160,43){{$\mu^-$}}
\put(160, 0){{$\Pe^-$}}
\SetScale{1}
\end{picture}
}
}
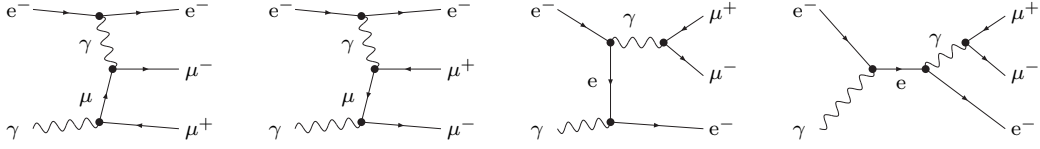
\caption{QED diagrams contributing to $\Pem\ga\to\Pem\mu^-\mu^+$ at
tree level.}
\label{fig:diags}
\efi
The corresponding helicity amplitudes,
including the full dependence on the masses $\Me$ and $m_\mu$, can be
obtained from the treatment of $\Pem\ga\to\Pem\Pem\Pep$ presented
in \citere{Denner:1998nk} after some obvious substitutions.
In the following we compare the result with the full mass dependence
to results obtained with the described subtraction and slicing methods
in various kinematical situations. Denoting the polar angle of
an outgoing particle $i$ by $\theta_i$ and the
angle between the two outgoing muons 
by $\alpha_{\mu\mu}$, we distinguish the following cases:
\renewcommand{\labelenumi}{\alph{enumi})}
\begin{enumerate}
\item
{\it No collinear splittings} \\
Angular cuts:
$\theta_{\cut}<\theta_\Pem <180^\circ-\theta_{\cut}$ and 
$\theta_{\mu^\pm} <180^\circ-\theta_{\cut}$ and 
$\theta_{\cut}<\alpha_{\mu\mu}$. \\
No collinear singularities are included, and the integrated
cross section is well defined for vanishing fermion masses, i.e.\
none of the subtraction methods has to be applied.
The difference between massive and massless calculations indicates
the size of the fermion mass effects.
\item
{\it Collinear splitting $\ga\to \Pem\Pep^*$} \\
Angular cuts:
$\theta_{\cut}<\theta_\Pem$ and 
$\theta_{\mu^\pm} <180^\circ-\theta_{\cut}$ and 
$\theta_{\cut}<\alpha_{\mu\mu}$. \\
The collinear splitting $\ga\to \Pem\Pep^*$ of the incoming photon
is integrated over, so that the third diagram of \reffi{fig:diags}
develops a collinear singularity for backward-scattered electrons.
The methods of \refse{se:affst} are applied
to the calculation with massless fermions.
\item
{\it Collinear splittings $\ga\to\mu^\mp\mu^{\pm*}$} \\
Angular cuts:
$\theta_{\cut}<\theta_\Pem <180^\circ-\theta_{\cut}$ and
$\theta_{\cut}<\alpha_{\mu\mu}$. \\
The collinear splittings $\ga\to\mu^\mp\mu^{\pm*}$
of the incoming photon are integrated over, 
so that the first two diagrams of \reffi{fig:diags}
develop collinear singularities for backward-scattered muons.
The methods of \refse{se:affst} are applied
to the calculation with massless fermions.
\item
{\it Collinear splitting $\ga^*\to\mu^-\mu^+$} \\
Angular cuts:
$\theta_{\cut}<\theta_\Pem <180^\circ-\theta_{\cut}$ and 
$\theta_{\mu^\pm} <180^\circ-\theta_{\cut}$. \\
The collinear splitting $\ga^*\to\mu^-\mu^+$
of an intermediate photon is integrated over, 
so that the last two diagrams of \reffi{fig:diags}
develop collinear singularities for collinearly produced muons.
The methods of \refse{se:astff} are applied
to the calculation with massless fermions.
\item
{\it Collinear splitting $\Pem\to\Pem\ga^*$} \\
Angular cuts:
$\theta_\Pem <180^\circ-\theta_{\cut}$ and 
$\theta_{\mu^\pm}<180^\circ-\theta_{\cut}$ and 
$\theta_{\cut}<\alpha_{\mu\mu}$. \\
The collinear splitting $\Pem\to\Pem\ga^*$ of the incoming electron
is integrated over, 
so that the first two diagrams of \reffi{fig:diags}
develop collinear singularities for forward-scattered electrons.
The methods of \refse{se:ffast} are applied
to the calculation with massless fermions.
\end{enumerate}
For the numerical evaluation we set 
the fermion masses to $\Me=0.51099907\MeV$ and $m_\mu=0.10565839\GeV$, 
the fine-structure constant to $\alpha=e^2/(4\pi)=1/137.0359895$,
the beam energies to $E=E_\Pe=E_\ga=250\GeV$,
and the angular cut to $\theta_{\cut}=10^\circ$. 
In the subtraction and slicing methods the masses $\Me$ and $m_\mu$
are neglected everywhere except for the mass-singular logarithms, i.e.\
the laboratory frame defined by the above beam energies coincides
with the centre-of-mass system. For the fully massive calculation the
two frames are connected by a (numerically irrelevant) boost along the 
beam axis with a tiny boost velocity of ${\cal O}(\Me^2/E^2)$.
Our numerical results for the different kinematical situations and
the various methods are collected in \refta{tab:results}.
\btab
\centerline{
\begin{tabular}{llll}
Collinear splittings & Method & $\sigma_{+-}[\pb]$ & $\sigma_{++}[\pb]$ 
\\
\hline
a) none & full mass dependence  & 0.50910(9)  & 0.47172(6) \\
        & massless case         & 0.51110(9)  & 0.47384(7) \\
\hline
b) $\ga\to \Pem\Pep^*$ &
full mass dependence            & 0.52213(7)  & 0.56762(7) \\
& subtraction (IS spectator)    & 0.52424(8)  & 0.57027(8) \\
& subtraction (FS spectator)    & 0.52434(7)  & 0.57017(9) \\
& slicing ($\De\theta=10^{-1}$) & 0.52410(7)  & 0.57021(6) \\
%& slicing ($\De\theta=10^{-2}$) & 0.52430(8)  & 0.57016(7) \\
& slicing ($\De\theta=10^{-3}$) & 0.52431(9)  & 0.57021(7) \\
%& slicing ($\De\theta=10^{-4}$) & 0.52427(7)  & 0.57028(7) \\
& slicing ($\De\theta=10^{-5}$) & 0.52423(8)  & 0.57028(7) \\
\hline
c) $\ga\to\mu^\mp\mu^{\pm*}$ &
full mass dependence            & 2.5890(5)   & 2.3615(4)  \\
& subtraction (IS spectator)    & 2.5872(3)   & 2.3586(5)  \\
& subtraction (FS spectator)    & 2.5873(8)   & 2.3585(5)  \\
& slicing ($\De\theta=10^{-1}$) & 2.5883(3)   & 2.3609(2)  \\
%& slicing ($\De\theta=10^{-2}$) & 2.5863(5)   & 2.3585(5)  \\
& slicing ($\De\theta=10^{-3}$) & 2.5859(8)   & 2.3578(8)  \\
%& slicing ($\De\theta=10^{-4}$) & 2.5857(11)  & 2.3590(10) \\
& slicing ($\De\theta=10^{-5}$) & 2.5876(13)  & 2.3572(13) \\
\hline
d) $\ga^*\to\mu^-\mu^+$ &
full mass dependence            & 0.54076(8)  & 0.53357(8)  \\
& subtraction (IS spectator)    & 0.54309(8)  & 0.53597(7)  \\
& subtraction (FS spectator)    & 0.54306(8)  & 0.53603(7)  \\
& slicing ($\De\theta=10^{-1}$) & 0.53164(19) & 0.52386(16) \\
%& slicing ($\De\theta=10^{-2}$) & 0.54283(19) & 0.53567(14) \\
& slicing ($\De\theta=10^{-3}$) & 0.54287(17) & 0.53624(15) \\
%& slicing ($\De\theta=10^{-4}$) & 0.54308(21) & 0.53581(15) \\
& slicing ($\De\theta=10^{-5}$) & 0.54335(18) & 0.53580(18) \\
\hline
e) $\Pem\to\Pem\ga^*$ &
full mass dependence            & 5.5465(7)   & 4.7060(6) \\
& subtraction (IS spectator)    & 5.5495(4)   & 4.7070(3) \\
& subtraction (FS spectator)    & 5.5484(6)   & 4.7064(5) \\
& slicing ($\De\theta=10^{-1}$) & 5.5313(1)   & 4.6880(1) \\
%& slicing ($\De\theta=10^{-2}$) & 5.5491(2)   & 4.7067(2) \\
& slicing ($\De\theta=10^{-3}$) & 5.5488(3)   & 4.7071(3) \\
%& slicing ($\De\theta=10^{-4}$) & 5.5490(4)   & 4.7067(4) \\
& slicing ($\De\theta=10^{-5}$) & 5.5486(5)   & 4.7067(4) \\
\end{tabular}
}
\caption{QED cross sections $\sigma_{\kappa\la}$ for
$\Pem\ga\to\Pem\mu^-\mu^+$ in the various setups described in the
main text, with signs $\kappa$ and $\la$ of the helicities of the incoming
$\Pem$ and $\ga$, respectively. The results are obtained with the
indicated methods, where IS and FS stand for spectators in the
initial and final states, respectively.}
\label{tab:results}
\etab
In addition in \refta{tab:results2}
we show the analogous results for the situation where the energy of each
final-state lepton $l=\Pe^-,\mu^\pm$ is restricted by $E_l>10\GeV$.
\btab
\centerline{
\begin{tabular}{llll}
Collinear splittings & Method & $\sigma_{+-}[\pb]$ & $\sigma_{++}[\pb]$ 
\\
\hline
a) none & full mass dependence  & 0.45780(6)  & 0.41699(6) \\
        & massless case         & 0.45779(6)  & 0.41704(5) \\
\hline
b) $\ga\to \Pem\Pep^*$ &
full mass dependence            & 0.46995(6)  & 0.50351(6) \\
& subtraction (IS spectator)    & 0.46999(6)  & 0.50345(6) \\
& subtraction (FS spectator)    & 0.46995(6)  & 0.50348(7) \\
& slicing ($\De\theta=10^{-1}$) & 0.46990(7)  & 0.50349(5) \\
%& slicing ($\De\theta=10^{-2}$) & 0.46997(7)  & 0.50351(7) \\
& slicing ($\De\theta=10^{-3}$) & 0.46992(7)  & 0.50352(5) \\
%& slicing ($\De\theta=10^{-4}$) & 0.46991(8)  & 0.50353(6) \\
& slicing ($\De\theta=10^{-5}$) & 0.46992(7)  & 0.50355(6) \\
\hline
c) $\ga\to\mu^\mp\mu^{\pm*}$ &
full mass dependence            & 2.4934(5)   & 2.2637(4)  \\
& subtraction (IS spectator)    & 2.4931(3)   & 2.2637(2)  \\
& subtraction (FS spectator)    & 2.4923(6)   & 2.2642(5)  \\
& slicing ($\De\theta=10^{-1}$) & 2.4895(2)   & 2.2606(2)  \\
%& slicing ($\De\theta=10^{-2}$) & 2.4921(5)   & 2.2634(5)  \\
& slicing ($\De\theta=10^{-3}$) & 2.4917(7)   & 2.2628(7)  \\
%& slicing ($\De\theta=10^{-4}$) & 2.4913(10)  & 2.2639(9)  \\
& slicing ($\De\theta=10^{-5}$) & 2.4905(12)  & 2.2626(12) \\
\hline
d) $\ga^*\to\mu^-\mu^+$ &
full mass dependence            & 0.48606(7)  & 0.47396(8)  \\
& subtraction (IS spectator)    & 0.48620(7)  & 0.47407(6)  \\
& subtraction (FS spectator)    & 0.48630(6)  & 0.47401(6)  \\
& slicing ($\De\theta=10^{-1}$) & 0.47588(19) & 0.46363(13) \\
%& slicing ($\De\theta=10^{-2}$) & 0.48570(19) & 0.47402(16) \\
& slicing ($\De\theta=10^{-3}$) & 0.48607(19) & 0.47399(14) \\
%& slicing ($\De\theta=10^{-4}$) & 0.48621(16) & 0.47405(20) \\
& slicing ($\De\theta=10^{-5}$) & 0.48623(20) & 0.47425(15) \\
\hline
e) $\Pem\to\Pem\ga^*$ &
full mass dependence            & 5.4878(6)   & 4.6467(5) \\
& subtraction (IS spectator)    & 5.4866(3)   & 4.6471(3) \\
& subtraction (FS spectator)    & 5.4871(5)   & 4.6475(5) \\
& slicing ($\De\theta=10^{-1}$) & 5.4690(1)   & 4.6278(1) \\
%& slicing ($\De\theta=10^{-2}$) & 5.4863(2)   & 4.6463(2) \\
& slicing ($\De\theta=10^{-3}$) & 5.4869(3)   & 4.6467(3) \\
%& slicing ($\De\theta=10^{-4}$) & 5.4866(4)   & 4.6469(3) \\
& slicing ($\De\theta=10^{-5}$) & 5.4862(5)   & 4.6466(4) \\
\end{tabular}
}
\caption{Same as in \refta{tab:results}, but with an energy cut
$E_l>10\GeV$ for all final-state leptons.}
\label{tab:results2}
\etab
All results are obtained with an integration by {\sl Vegas}
\cite{Lepage:1977sw}, using $25\times 10^6$ events. While a simple phase-space
parametrization is sufficient in the subtraction formalism, 
dedicated phase-space mappings are required to flatten the
corresponding collinear poles in the slicing approach and when
employing the full mass dependence of the matrix elements.
The fully massive results have been checked with the program
{\sc Whizard}~\cite{Kilian:2007gr}, where agreement within the integration
errors has been found.

The results obtained with the different subtraction variants,
where a spectator is chosen from the initial state (IS) or from the
final state (FS), are in mutual agreement within the integration error,
which is indicated in parentheses. Subtraction and slicing results
are also consistent within the statistical errors as long as the
angular slicing cut $\De\theta$ is not chosen too large.
For example, some of the slicing results for $\De\theta=10^{-1}$
still show a significant residual dependence on $\De\theta$.
In the chosen example,
the integration errors of the subtraction and slicing results are of
the same order of magnitude. However, we would like to mention that
the subtraction approach is often more efficient, as e.g.\ observed
in the applications of 
\citeres{Bredenstein:2006rh,Dittmaier:2001ay,%
Brensing:2007qm,Diener:2003ss,Ciccolini:2007jr} mentioned above.
This superiority of the subtraction formalism typically deteriorates
if complicated phase-space cuts are applied, as in the chosen example,
because the cuts act differently in the various auxiliary phase
spaces and thus introduce new peak structures in the integrand.

Finally, we remark that the impact of mass-suppressed terms is 
significantly reduced if the cut on the lepton energies $E_l$ is
applied. This cut guarantees that $E_l\gg m_l$ overall in phase space,
so that mass-suppressed terms are proportional 
to $m_l^2/Q^2$ with $Q\gg m_l$. Without any restriction on $E_l$,
there are at least small regions of phase space where $Q$ is not much
smaller than $m_l$, leading to larger mass effects. This feature is
clearly visible in \reftas{tab:results} and \ref{tab:results2}
when comparing results based on the full mass 
dependence in the matrix elements with the subtraction and slicing
results that are based on the asymptotic limit $m_l\to0$.

\section{Summary}
\label{se:summary}

The dipole subtraction formalism for photonic
corrections is extended to various photon--fermion splittings
where the resulting collinear singularities lead to corrections
that are enhanced by logarithms of small fermion masses $m_f$.
Specifically, we have considered non-collinear-safe final-state radiation,
collinear fermion production from incoming photons, forward-scattered
incoming fermions, and collinearly produced fermion--antifermion pairs.
All formulas needed in applications are provided, only the scattering
matrix elements for the underlying process and for relevant subprocesses
have to be supplemented in the simple approximation of a massless
fermion $f$. Particle polarization is taken care of in all
relevant cases, e.g., for incoming fermions and photons.
For the purpose of cross-checking results in applications, we
also provide the formulas needed in the phase-space slicing method.

As an example illustrating the use and performance of the proposed methods
we have explicitly applied the subtraction procedures to the process
$\Pem\ga\to\Pem\mu^-\mu^+$ and compared the results to those
obtained with phase-space slicing. 
The presented subtraction variants will certainly be used in
several precision calculations needed for present and future
collider experiments such as the LHC or ILC.

\section*{Acknowledgement}

We are grateful to Markus Roth for collaborating on the issue of 
\refse{se:fsr} in an early stage of this work.

\appendix
\section*{Appendix}

\section{More details on non-collinear-safe final-state radiation}
\label{app:fsr}

Here we generalize the results of \refse{se:noncollsafe}, where
non-collinear-safe photon radiation off fermions is treated,
to the situation where massive spectators in the final state exist.
To this end, we only have to consider the case of final-state emitter 
and final-state spectator.

For $m_i\to0$, $m_\ga=0$, but $m_j\ne0$, the boundary of the $y_{ij}$ 
integration [given for the massless case in Eq.~\refeq{eq:yboundary}]
is given by
\beqar
y_1(z) &=& \frac{m_i^2(1-z)}{\bar P_{ij}^2 z}, 
\nn\\
y_2(z) &=& \left[\xi(z)+1+\sqrt{\xi(z)[\xi(z)+2]}\right]^{-1} \quad
\mbox{with} \quad \xi(z)=\frac{m_j^2}{2\bar P_{ij}^2 z(1-z)},
\eeqar
and the functions relevant for the integrand $\gsub_{ij,\tau}$ behave as
\beqar
p_i k &=& \frac{\bar P_{ij}^2 y_{ij}}{2}, \qquad
\bar P_{ij}^2 = P_{ij}^2-m_j^2, 
\nn\\
R_{ij}(y) &=& \sqrt{(1-y)^2-\frac{4m_j^2 y}{\bar P_{ij}^2}}, \qquad 
r_{ij}(y) = 1-\frac{2m_j^2}{\bar P_{ij}^2}\frac{y}{1-y}.
\eeqar
The evaluation of Eq.~\refeq{eq:bcGsubij} now becomes non-trivial
and yields
\beqar
\bcGsub_{ij,+}(P_{ij}^2,z) &=&
-P_{ff}(z)\ln\biggl(\frac{m_i^2}{z\bar P_{ij}^2}[1-\eta(z)]\biggr)
+(1+z)\ln(1-z)
-\frac{2z}{1-z}
\nn\\
&& {}
+(1+z)\ln\biggl(1+\frac{m_j^2}{\bar P_{ij}^2\eta(z)}\biggr)
-\frac{2}{(1-z)\si(z)} \Biggl\{
\ln\biggl(1+\frac{\bar P_{ij}^2\eta(z)[1-z\eta(z)]}{m_j^2(1-z)}\biggr)
\nn\\
&& \qquad {}
-2\ln\biggl(1-\frac{2z\eta(z)}{1+\si(z)}\biggr)
+\si(z)\ln\biggl(\frac{m_j^2}{\bar P_{ij}^2\eta(z)}(1-z)\biggr)
\Biggr\}
-\bcGsub_{ij,-}(P_{ij}^2,z),
\nn\\[.5em] 
\bcGsub_{ij,-}(P_{ij}^2,z) &=& 1-z,
\label{eq:bcGsubij3}
\eeqar
with the auxiliary functions
\beq
\si(z) = \sqrt{1 + \frac{4m_j^2}{\bar P_{ij}^2}z(1-z)}, \qquad
\eta(z) = \left\{
\barr{ll}   [1-y_2(z)]z     & \mbox{ for } z<\frac{1}{2}, \\ 
          {}[1-y_2(z)](1-z) & \mbox{ for } z>\frac{1}{2}. \earr\right.
\eeq
For $m_j\to0$, the results for $\bcGsub_{ij,\tau}(P_{ij}^2,z_{ij})$
reduce to Eq.~\refeq{eq:bcGsubij2}, as can be easily seen after realizing
that $\eta(z)={\cal O}(m_j)$ and $\si(z)=1+{\cal O}(m_j^2)$ in this limit.

\section{\boldmath{More details on the subtraction for 
$\ga\to f\bar f^*$ splittings}}

\subsection{Factorization in the collinear limit}
\label{app:affst-fac}

In this section we derive the asymptotic behaviour \refeq{eq:affst-fact}
of the squared
amplitude $|{\cal M}_{\gamma a\to fX}|^2$ for the case where the
outgoing light fermion flies along the direction of the incoming photon.
We consider polarized incoming photons with momentum $k^\mu$ and 
polarization vector $\veps^\mu_{\la_\ga}$, where
$\la_\ga=\pm$ is the sign of its helicity. We further introduce a light-like
gauge vector $n^\mu$ ($n^2=0, nk\ne0$), i.e.\ $\veps^\mu_{\la_\ga}$
is characterized by
\beq
\veps^\mu_{-\la_\ga} = (\veps^\mu_{\la_\ga})^*, \qquad
k\veps_{\la_\ga} = n\veps_{\la_\ga} = 0.
\eeq
In the following we make use of the identity%
\footnote{This identity is easily proven using a representation of the
polarization vectors by Weyl spinors. Employing the conventions
of \citere{Dittmaier:1998nn}, we have 
$\veps^{\dot AB}_+ = \veps^\mu_+ \sigma^{\dot AB}_\mu =
\sqrt{2} n^{\dot A}k^B/\langle kn\rangle$ and
$\veps^{\dot AB}_- = \veps^\mu_- \sigma^{\dot AB}_\mu =
\sqrt{2} k^{\dot A}n^B/\langle kn\rangle^*$ for the polarization
bispinors, so that \\
$\begin{array}{rcl}
\eps^{\mu\nu\rho\sigma}k_\rho n_\sigma 
&=& \frac{\ri}{4}( \eps^{\dot A\dot E}\eps^{\dot C\dot G}\eps^{BD}\eps^{FH}
-\eps^{\dot A\dot C}\eps^{\dot E\dot G}\eps^{BF}\eps^{DH} )
\sigma^\mu_{\dot AB} \sigma^\nu_{\dot CD} 
\sigma^\rho_{\dot EF} \sigma^\sigma_{\dot GH} k_\rho n_\sigma \\
&=& \frac{\ri}{4}( k^{\dot A} n^{\dot C}\eps^{BD}k_X n^X
-\eps^{\dot A\dot C}k^B n^D k_{\dot X}n^{\dot X} )
\sigma^\mu_{\dot AB} \sigma^\nu_{\dot CD} \\
&=& \frac{\ri}{4}( -k^{\dot A}n^B n^{\dot C} k^D 
+ n^{\dot A} k^B k^{\dot C} n^D ) \sigma^\mu_{\dot AB} \sigma^\nu_{\dot CD} \\
&=& \frac{\ri}{2}\langle kn\rangle\langle kn\rangle^*
(\veps^\mu_+ \veps^\nu_- - \veps^\mu_- \veps^\nu_+) 
= \ri(kn)(\veps^\mu_+ \veps^\nu_- - \veps^\mu_- \veps^\nu_+). 
\earr$ \\
The only non-trivial step is the third equality which follows from a
twofold application of Schouten's identity.
}
\beq
\veps^\mu_\pm (\veps^\nu_\pm)^* = \veps^\mu_\pm \veps^\nu_\mp = 
\frac{1}{2}E^{\mu\nu}(k) 
\mp\frac{\ri}{2kn}\eps^{\mu\nu\rho\sigma}k_\rho n_\sigma,
\label{eq:apolid}
\eeq
where
\beq
E^{\mu\nu}(k) = \veps^\mu_+ \veps^\nu_- +\veps^\mu_- \veps^\nu_+
= -g^{\mu\nu}+\frac{k^\mu n^\nu+n^\mu k^\nu}{kn}
\eeq
is the polarization sum of the photon in four space-time dimensions
and $\eps^{\mu\nu\rho\sigma}$ the Levi-Civita tensor with $\eps^{0123}=+1$.

In a gauge for the photon where $nk={\cal O}(k^0)$, it is
easily shown by power counting that the logarithmic singularity 
arising from the phase-space region $k p_f={\cal O}(m_f^2)$
($m_f\ll k^0$) originates from diagrams in which
the incoming photon collinearly splits into a light $f\bar f^*$ pair.
The generic form of such graphs is shown in \reffi{fig:affst2}.
\bfi
\centerline{ 
\begin{picture}(450,120)(0,0)
\put(0,-5){ 
  \begin{picture}(250,120)(0,0)
  \Photon(15,70)(50,70){2}{4}
  \Photon(245,70)(210,70){2}{4}
  \ArrowLine(90,40)(50,70)
  \ArrowLine(210,70)(170,40)
  \Line(90,40)( 50, 10)
  \Line(170,40)(210, 10)
  \LongArrow( 70,88)( 85, 93)
  \LongArrow( 20,77)( 35, 77)
  \LongArrow( 55,57)( 70, 46)
  \Vertex( 50,70){2.5}
  \Vertex(210,70){2.5}
  \ArrowArc(130,-57)(-150,-90,-57)
  \ArrowArc(130,-57)(-150,-123,-90)
  \CArc(130,10)(50,45,135)
  \CArc(130,70)(50,-135,-45)
  \GCirc( 90,40){10}{1}
  \GCirc(170,40){10}{1}
  \DashLine(130,5)(130,120){6}
  \put(  3, 68){$\gamma$}
  \put( 85, 75){$f$}
  \put( 72, 60){$\bar f$}
  \put( 25, 83){$k$}
  \put( 72, 98){$p_f$}
  \put( 50, 42){$p_{\bar f}$}
  \put( 40,  2){$a$}
  \put(108, 36){\huge$\vdots$\normalsize $X$}
  \put( 78,115){${\cal M}_{\gamma a\to fX}$}
  \put(140,115){${\cal M}_{\gamma a\to fX}^*$}
  \end{picture} } 
\put(240,-5){ 
  \begin{picture}(210,120)(0,0)
  \ArrowLine(90,40)(50,70)
  \ArrowLine(210,70)(170,40)
  \Line(90,40)( 50, 10)
  \Line(170,40)(210, 10)
  \LongArrow( 55,57)( 70, 46)
  \CArc(130,10)(50,45,135)
  \CArc(130,70)(50,-135,-45)
  \GCirc( 90,40){10}{1}
  \GCirc(170,40){10}{1}
  \DashLine(130,5)(130,95){6}
  \put( 72, 60){$\bar f$}
  \put( 50, 42){$p_{\bar f}$}
  \put( 40,  2){$a$}
  \put(108, 36){\huge$\vdots$\normalsize $\,X$}
  \put( 80,90){${\cal M}_{\bar fa\to X}$}
  \put(140,90){${\cal M}_{\bar fa\to X}^*$}
  \end{picture} } 
\end{picture} }
\caption{Generic squared diagram for the splitting $\ga\to f\bar f^*$ 
(left) and the corresponding squared diagram for the related process
with an incoming $\bar f$ (right).}
\label{fig:affst2}
\efi
Assuming summation over the polarization of the outgoing fermion $f$,
the squared matrix element, thus, behaves like
\beqar
\lefteqn{
|\M_{\gamma a\to fX}(k,p_a,p_f;\la_\ga)|^2}
\nn\\
&\asymp{kp_f\to0}&
Q_f^2 e^2 \, \bar T_{\bar fa\to X}(p_{\bar f},p_a) 
\frac{-\dsl{p}_{\bar f}+m_f}{p_{\bar f}^2-m_f^2}
\dsl{\veps}^*_{\la_\ga} (\dsl{p}_f+m_f) \dsl{\veps}_{\la_\ga}
\frac{-\dsl{p}_{\bar f}+m_f}{p_{\bar f}^2-m_f^2} 
T_{\bar fa\to X}(p_{\bar f},p_a),
\label{eq:M2affst}
\eeqar
where $T_{\bar fa\to X}(p_{\bar f},p_a)$ includes all information of the 
subamplitude indicated by the open blob in \reffi{fig:affst2} and
$\bar T_{\bar fa\to X} = (T_{\bar fa\to X})^\dagger\ga_0$.
To leading order in $m_f\to0$,
the squared amplitude for the subprocess $\bar fa\to X$ can
be written as
\beq
|\M_{\bar fa\to X}(p_{\bar f},p_a;\kappa_{\bar f})|^2
= \bar T_{\bar fa\to X}(p_{\bar f},p_a)
\, \Bigl[ \omega_{-\kappa_{\bar f}} \dsl{p}_{\bar f}
+{\cal O}(m_f)\Bigr] \,
T_{\bar fa\to X}(p_{\bar f},p_a),
\eeq
with $\omega_\pm=\frac{1}{2}(1\pm\gamma_5)$ and $\kappa_{\bar f}=\pm$
denoting the sign of the $\bar f$ helicity. 
In order to find the relation between these squared matrix elements,
we insert identity \refeq{eq:apolid} into Eq.~\refeq{eq:M2affst}
and eliminate the $\eps$~tensor via Chisholm's identity,
\beq
\ri\epsilon^{\al\be\ga\de}\ga_\de = (\ga^\al \ga^\be \ga^\ga
- g^{\al\be}\ga^\ga + g^{\al\ga}\ga^\be - g^{\be\ga}\ga^\al)\ga_5,
\eeq
i.e.\ we trade the $\eps$~contributions for a $\ga_5$ insertion
in the spinor chain.
Next, we isolate the leading terms in the collinear limit
$kp_f={\cal O}(m_f^2)\to0$. This limit can, e.g., be
parametrized by the decomposition of the momentum of $f$
\beq
p_f^\mu = (1-x)k^\mu + p_{f,\perp}^\mu + p_{f,r}^\mu
\eeq
with $x=1-p_f^0/k^0$, 
$kp_{f,\perp}=0$, and ${\bf p}_{f,r}={\bf 0}$ 
(where boldface symbols refer the spatial parts of momenta).
In this decomposition we have 
${\cal O}(p_{f,\perp}^0)={\cal O}(p_{f,r}^0)= {\cal O}(m_f^2)$ and
${\bf p}_{f,\perp}^2={\cal O}(m_f^2)$.
Thus, each component of the orthogonal 3-vector ${\bf p}_{f,\perp}$
is of ${\cal O}(m_f)$. 
After some straightforward simplifying algebra,
the result of applying the power counting to $|\M_{\gamma a\to fX}|^2$ is%
\footnote{Actually there are also terms proportional to
$\la_\ga\,m_f/(kp_f)^2\,\bar T_{\bar fa\to X} \dsl{k} \dsl{p}_{f,\perp}\ga_5
T_{\bar fa\to X}$, which at first sight seem to 
contribute in ${\cal O}(m_f^{-2})$
in the limit $m_f\to0$. Although these terms obviously disappear from the
subtraction function after setting $m_f$ to zero, they potentially
contribute to the corresponding integrated subtraction terms, in which
the limit $m_f\to0$ is taken after the singular phase-space integration.
However, the integration over the
azimuthal angle of ${\bf p}_f$, which is always assumed
in our analysis, leads to a further suppression by one power of $m_f$,
so that the contribution to the phase-space integral of 
$|\M_{\gamma a\to fX}|^2$ is mass suppressed.
Thus, these terms are irrelevant in the construction of a
subtraction function to separate mass-singular terms in the collinear cone.}
\beqar
\lefteqn{
|\M_{\gamma a\to fX}(k,p_a,p_f;\la_\ga)|^2}
\nn\\
&\asymp{kp_f\to0}& \frac{Q_f^2e^2}{2x(kp_f)} \,
\bar T_{\bar fa\to X}(p_{\bar f},p_a) \, \Biggl\{
\biggl[1-2x(1-x)+\frac{xm_f^2}{kp_f}\biggr] \dsl{p}_{\bar f}
\nn\\
&& \hspace*{3em} {}
- \la_\ga \biggl[2x-1+\frac{m_f^2}{kp_f}\biggr] \ga_5 \dsl{p}_{\bar f}
\Biggr\} \,T_{\bar fa\to X}(p_{\bar f},p_a)
\nn\\
&\asymp{kp_f\to0}&
\frac{1}{2}Q_f^2e^2 \, \Biggl\{
(h^{\ga f}_+ + h^{\ga f}_-)
\left[ |\M_{\bar fa\to X}(p_{\bar f},p_a;+)|^2
      +|\M_{\bar fa\to X}(p_{\bar f},p_a;-)|^2 \right]
\nn\\
&& \hspace*{3em} {}
+ \la_\ga (h^{\ga f}_+ - h^{\ga f}_-)
\left[ |\M_{\bar fa\to X}(p_{\bar f},p_a;+)|^2
      -|\M_{\bar fa\to X}(p_{\bar f},p_a;-)|^2 \right]
\Biggr\}. 
\hspace{2em}
\eeqar
The last form results from the last but one by simply
substituting $|\M_{\bar fa\to X}|^2$ and the $h^{\ga f}_\pm$ functions
defined in Eq.~\refeq{eq:haf}, whose arguments are suppressed in the
notation. This completes our proof of Eq.~\refeq{eq:affst-fact},
which is a more compact version of this result.

In this section we have explicitly treated $f$ as fermion and $\bar f$
as antifermion. The opposite case with $f$ being an antifermion and
$\bar f$ a fermion is obtained analogously and leads to the identical 
final result \refeq{eq:affst-fact}, although some signs in intermediate
results are different. This fact is, of course, to be expected, because
relations between squared helicity amplitudes cannot depend on our
convention which fermion we call the antiparticle of the other.

\subsection{\boldmath{Dipole subtraction for $\ga\to f\bar f^*$ splittings
with massive final-state spectator}}
\label{app:affst-fin}

Here we give some details on the derivation of the integrated subtraction
part presented in \refse{se:affst-fin}
for the collinear splitting $\ga\to f\bar f^*$ in the process
$\gamma a\to fjX$, where $j$ is a possibly massive spectator.
We start by generalizing the form \refeq{eq:momaffst-fin} of the new
momenta upon restoring the correct dependence on $m_f$,
\beqar
\tilde p_{\bar f}^\mu(x) &=& \frac{\sqrt{\la_{fj,\ga}}}{-\bar P^2}
\left( x k^\mu + \frac{\bar P^2}{2P^2}P^\mu
\right) - \frac{P^2+m_f^2-m_j^2}{2P^2}P^\mu,
\qquad
\tilde p_{\bar f}^\mu = \tilde p_{\bar f}^\mu(x_{fj,\ga}),
\nn\\
\tilde p_j^\mu(x) &=& P^\mu + \tilde p_{\bar f}^\mu(x), \qquad
\tilde p_j^\mu = \tilde p_j^\mu(x_{fj,\ga}),
\label{eq:momaffst-fin2}
\eeqar
where the following shorthands are used,
\beq
\bar P^2 = P^2-m_f^2-m_j^2, \quad
\la_{fj,\ga} = \la(P^2,m_f^2,m_j^2),
\eeq
with the auxiliary function
\beq
\lambda(x,y,z) = x^2+y^2+z^2-2xy-2xz-2yz.
\eeq
The new momenta satisfy the on-shell conditions $\tilde p_{\bar f}^2=m_f^2$,
$\tilde p_j=m_j^2$ and correctly behave in the collinear limit,
$\tilde p_{\bar f} \to x k$, where $k p_f={\cal O}(m_f^2)\to0$.
The splitting of the $(N+1)$-particle phase space into the corresponding
$N$-particle phase space and the integral over the remaining
singular degrees of freedom is given by
\beq
\int\rd\phi(p_f,p_j,k_X;k+p_a) 
= \int_{0}^{x_1}\rd x
\int\rd\phi\Big(\tilde p_j(x),k_X;\tilde p_{\bar f}(x)+p_a\Big)
\int [\rd p_f(P^2,x,z_{fj,\ga})],
\eeq
with the explicit parametrization 
\beq
\int [\rd p_f(P^2,x,z_{fj,\ga})] = \frac{1}{2(2\pi)^3} 
\frac{-\bar P^2(p_a\tilde p_{\bar f}(x))}{x^2s}
\int_{z_1(x)}^{z_2(x)} \rd z_{fj,\ga} \,
\int \rd\phi_f,
\eeq
and $k_X$ denoting the outgoing total momentum of $X$.
The upper kinematical limit of the parameter $x=x_{fj,\ga}$ is given by
\beq
x_1 = \frac{-\bar P^2}{-\bar P^2+2m_f m_j}.
\eeq
The integration of the azimuthal angle $\phi_f$ of $f$ simply
yields a factor $2\pi$, but the integration of the auxiliary parameter
\beq
z_{fj,\ga} = \frac{k p_j}{k p_f+k p_j}
\eeq
with the boundary
\beq
z_{1,2}(x) = \frac{2m_j^2 x+\bar P^2(x-1) \mp
	\sqrt{\bar P^4(1-x)^2-4m_f^2 m_j^2 x^2}}%
		{2(P^2 x-\bar P^2)}
\eeq
is non-trivial.
The integration kernels occurring in the final result \refeq{eq:sigaffst-fin}
are defined as
\beq
\cHsub^{\ga f}_{j,\tau}(P^2,x) = \frac{-\bar P^2}{2} 
\int_{z_1(x)}^{z_2(x)} \rd z_{fj,\ga} \, 
\hsub^{\ga f}_{j,\tau}(k,p_f,p_j)
\eeq
and can be evaluated without problems analytically (even for finite $m_f$)
yielding Eq.~\refeq{eq:cHaffst-fin} for $m_f\to0$.

\section{\boldmath{More details on the subtraction for
$\ga^*\to f\bar f$ splittings}}
\label{app:astff-fin}

In this appendix we supplement \refse{se:astff-fin}, where the subtraction
procedure for collinear $\ga^*\to f\bar f$ splittings has been described
for a final-state spectator $j$. In the following we fully take into
account the spectator mass $m_j$.
The derivation widely follows \citere{Catani:2002hc}, where the 
treatment of the $\Pg^*\to Q\bar Q$ splitting with a massive quark $Q$
has been considered. Our approach differs from the one described in 
\citere{Catani:2002hc} in the level of inclusiveness that is
assumed in the collinear limit; in contrast to that paper we do not
assume a recombination of the $f\bar f$ pair in the collinear limit, but
instead control the individual momentum flow of $f$ and $\bar f$.

For arbitrary mass values $m_f$ and $m_j$ the subtraction function
can be constructed as in Eq.~\refeq{eq:Msubastff-fin} with the
generalized radiator function
\beqar
h_{f\bar f,j}^{\mu\nu}(p_f,p_{\bar f},p_j) &=& 
\frac{2}{(p_f+p_{\bar f})^2v_j} \Biggl[-g^{\mu\nu}
\left(1-2\kappa\biggl(z_1 z_2-\frac{m_f^2}{(p_f+p_{\bar f})^2}\biggr)\right)
\nn\\
&& \hspace{6em} {}
-\frac{4}{(p_f+p_{\bar f})^2}
\Bigl(z_{f\bar fj}^{(m)} p^\mu_f-\bar z_{f\bar fj}^{(m)}p^\mu_{\bar f}\Bigr)
\Bigl(z_{f\bar fj}^{(m)} p^\nu_f-\bar z_{f\bar fj}^{(m)}p^\nu_{\bar f}\Bigr) \Biggr].
\hspace{2em}
\eeqar
In addition to the parameters $y_{f\bar fj}$ and $z_{f\bar fj}$, which are defined
as in Eq.~\refeq{eq:yzastff}, we make use of the following auxiliary
quantities,
\beqar
\bar P^2 &=& P^2-2m_f^2-m_j^2,
\nn\\
v_f &=& \sqrt{\frac{\bar P^2 y_{f\bar fj}-2m_f^2}{\bar P^2 y_{f\bar fj}+2m_f^2}},
\qquad
v_j = 
\frac{\sqrt{[2m_j^2+\bar P^2(1-y_{f\bar fj})]^2-4m_j^2 P^2}}
{\bar P^2(1-y_{f\bar fj})},
\nn\\
z_{1,2} &=& \frac{1}{2}(1\mp v_j v_f),
\qquad
z_{f\bar fj}^{(m)} = z_{f\bar fj}-\frac{1}{2}(1-v_j),
\qquad
\bar z_{f\bar fj}^{(m)} = \bar z_{f\bar fj}-\frac{1}{2}(1-v_j).
\label{eq:paramastfffin}
\eeqar
The parameter $\kappa$ is arbitrary, because the singular behaviour
does not depend on it; in practice the independence of the final
result on $\kappa$ can be used as check.
The auxiliary momenta entering the hard scattering matrix element
for the subprocess $ab\to\ga jX$ also become more complicated,
\beqar
\tilde p_j^\mu &=& 
\frac{P^2-m_j^2}{\sqrt{\la(P^2,(p_f+p_{\bar f})^2,m_j^2)}}
\biggl(p_j^\mu-\frac{Pp_j}{P^2}P^\mu\biggr)
+\frac{P^2+m_j^2}{2P^2} P^\mu,
\nn\\
\tilde k^\mu &=& P^\mu-\tilde p_j^\mu, \qquad 
P^\mu=p^\mu_f+p^\mu_{\bar f}+p_j^\mu.
\eeqar
In order to integrate the subtraction function we need the 
azimuthal-averaged version of $h_{f\bar f,j}^{\mu\nu}$,
\beqar
h_{f\bar f,j}(p_f,p_{\bar f},p_j) &=& 
\frac{2}{(p_f+p_{\bar f})^2v_j} \left[ P_{f\ga}(z_{f\bar fj})
+2(1-\kappa)z_1 z_2
+\frac{2\kappa m_f^2}{(p_f+p_{\bar f})^2}\right],
\hspace{2em}
\eeqar
and an appropriate splitting of the phase space of the momenta 
$p_f$, $p_{\bar f}$, $p_j$,
\beqar
\int\rd\phi(p_f,p_{\bar f},p_j;P) &=& 
\int\rd\phi(\tilde k,\tilde p_j;P)
\int [\rd p_f(P^2,y_{f\bar fj},z_{f\bar fj})],
\\
\int [\rd p_f(P^2,y_{f\bar fj},z_{f\bar fj})] &=&
\frac{1}{4(2\pi)^3} \, \frac{\bar P^4}{P^2-m_j^2} \,
\int_{y_1}^{y_2} \rd y_{f\bar fj} \, (1-y_{f\bar fj})\,
\int_{z_1(y_{f\bar fj})}^{z_2(y_{f\bar fj})} \rd z_{f\bar fj} \,
\int \rd\phi_f,
\nn
\eeqar
where
\beq
y_1 = \frac{2m_f^2}{\bar P^2}, \qquad
y_2 = 1-\frac{2m_j(\sqrt{P^2}-m_j)}{\bar P^2}
\eeq
and $z_{1,2}(y_{f\bar fj})$ are the $z_{1,2}$ of Eq.~\refeq{eq:paramastfffin},
evaluated as functions of $y_{f\bar fj}$.
Up to this point, the full dependence on $m_f$ and $m_j$ is kept.

Since we want to keep the momentum flow in the collinear limit open,
i.e.\ the $z_{f\bar fj}$ integration should be done numerically,
we have to interchange the order of $y_{f\bar fj}$ and $z_{f\bar fj}$
integrations in the singular
phase-space integration over $\int[\rd p_f]$.
For arbitrary masses $m_f$ and $m_j$, this seems hardly possible
analytically, so that we focus on the limit $m_f\to0$ in the following,
because this is the interesting case.
We define
\beqar
\cHsub_{f\bar f,j}(P^2,z) &=&
\frac{\bar P^2}{2} \,
\int_{y_1(z)}^{y_2(z)} \rd y_{f\bar fj} \, (1-y_{f\bar fj}) \, 
h_{f\bar f,j}(p_f,p_{\bar f},p_j), 
\nn\\
\Hsub_{f\bar f,j}(P^2) &=&
\int_0^1 \rd z\, \cHsub_{f\bar f,j}(P^2,z),
\eeqar
where we were allowed to use $m_f=0$ in the prefactors and in the 
integration limits of $z=z_{f\bar fj}$. The relevant asymptotics of
$y_{1,2}(z)$ for $m_f\to0$ is
\beq
y_1(z) = \frac{m_f^2}{\bar P^2}\,\frac{z^2+(1-z)^2}{z(1-z)},
\qquad
y_2(z) = \frac{\sqrt{4\bar P^2 z(1-z)+m_j^2}-m_j}%
	{\sqrt{4\bar P^2 z(1-z)+m_j^2}+m_j}.
\eeq
The actual integration over $y_{f\bar fj}$ yields
\beqar
\cHsub_{f\bar f,j}(P^2,z) &=& 
%\Bigl(z^2+(1-z)^2\Bigr) 
P_{f\ga}(z) \Biggl[
2\ln\left(\frac{\sqrt{4\bar P^2 z(1-z)+m_j^2}-m_j}{2m_f}\right)
-1-\eta(z)
\nn\\
&& {}
\hspace{6em}
-2\ln[1-\eta(z)]
+\frac{m_j^2[1-\eta(z)]}{m_j^2+\eta(z)\bar P^2} \, \Biggr]
\nn\\
&& {}
+\frac{2m_j^2}{\bar P^2}\Bigl(1-\kappa+z^2+(1-z)^2\Bigr)
\ln\biggl(1+\eta(z)\frac{\bar P^2}{m_j^2}\biggr) + 2z(1-z)
\eeqar
with
\beq
\eta(z) = \left\{
\barr{ll}   [1-y_2(z)]z     & \mbox{ for } z<\frac{1}{2}, \\ 
          {}[1-y_2(z)](1-z) & \mbox{ for } z>\frac{1}{2}. \earr\right.
\eeq
The case $m_j=0$ given in Eq.~\refeq{eq:Hastff} can be easily read off
after realizing that $\eta(z)={\cal O}(m_j)$.
For the evaluation of $\Hsub_{f\bar f,j}(P^2)$ it is easier to
integrate first over $z$ and then over $y_{f\bar fj}$. The result is
\beq
\Hsub_{f\bar f,j}(P^2) = 
\frac{4}{3}\ln\biggl(\frac{\sqrt{P^2}-m_j}{m_f}\biggr) - \frac{16}{9}
+\frac{4m_j}{3(\sqrt{P^2}+m_j)} 
+ \biggl(\kappa-\frac{2}{3}\biggr) \frac{2m_j^2}{\bar P^2}
\ln\biggl(\frac{2m_j}{\sqrt{P^2}+m_j}\biggr),
\eeq
which could also be derived from Eq.~(5.36) of \citere{Catani:2002hc}.
For $m_j=0$ this obviously leads to the form given in Eq.~\refeq{eq:Hastff}.

\section{\boldmath{More details on the subtraction for 
$f\to f\ga^*$ splittings}}

\subsection{Factorization in the collinear limit}
\label{app:ffast-fac}

In this section we derive the asymptotic behaviour \refeq{eq:ffast-fact}
of the squared
amplitude $|{\cal M}_{fa\to fX}|^2$ for the case where the incoming and
outgoing light fermions become collinear.
We consider polarized incoming fermions $f$ with momentum $p_f^\mu$ and 
helicity of sign $\kappa_f=\pm$.
The corresponding Dirac spinor $u(p_f,\kappa_f)$ is an eigenspinor
of the helicity projector 
\beq
\Sigma_{\kappa_f} = \frac{1}{2}(1+\kappa_f\gamma_5\dsl{s}_{p_f}),
\eeq
where the polarization vector 
\beq
s^\mu_{p_f} = 
\biggl(\frac{|{\bf p}_f|}{m_f},\frac{p_f^0}{m_f}{\bf e}_f\biggr)
\eeq
is aligned to the direction ${\bf e}_f={\bf p}_f/|{\bf p}_f|$ for
helicity eigenstates. Defining the light-like vectors
$\tilde k^\mu=k_0(1,{\bf e}_f)$ and 
$n^\mu=(1,-{\bf e}_f)$,
the polarization vector $s^\mu_{p_f}$ can be decomposed into
$\tilde k^\mu$ and $n^\mu$  as follows,
\beq
s^\mu_{p_f} = 
\frac{(p_f n)}{2m_f k_0} \tilde k^\mu - \frac{m_f}{2(p_f n)} n^\mu.
\label{eq:spfkn}
\eeq
Note that the momentum $k^\mu$ of the virtual photon fulfills
$kn={\cal O}(k^0)$ in the collinear limit, because then
$k^\mu = \tilde k^\mu + {\cal O}(m_f)$. The vector $n^\mu$
will be used as gauge vector in the explicit definition of
photon polarization vectors for the subprocess $\gamma a\to X$
below.

\begin{sloppypar}
Power counting reveals that the logarithmic singularity 
arising from the phase-space region $p_f p'_f={\cal O}(m_f^2)\to0$
($m_f\ll p_f^0$) originates from the square of diagrams in which
the incoming fermion collinearly emits a photon that triggers the
production of $X$.
The generic form of such graphs is shown in \reffi{fig:ffast2}.
\bfi
\centerline{ 
\begin{picture}(450,120)(0,0)
\put(0,-5){ 
  \begin{picture}(250,120)(0,0)
  \ArrowLine(15,70)(50,70)
  \ArrowLine(210,70)(245,70)
  \Photon(90,40)(50,70){2}{5}
  \Photon(210,70)(170,40){2}{5}
  \Line(90,40)( 50, 10)
  \Line(170,40)(210, 10)
  \LongArrow( 70,88)( 85, 93)
  \LongArrow( 20,77)( 35, 77)
  \LongArrow( 55,57)( 70, 46)
  \Vertex( 50,70){2.5}
  \Vertex(210,70){2.5}
  \ArrowArc(130,-57)(-150,-90,-57)
  \ArrowArc(130,-57)(-150,-123,-90)
  \CArc(130,10)(50,45,135)
  \CArc(130,70)(50,-135,-45)
  \GCirc( 90,40){10}{1}
  \GCirc(170,40){10}{1}
  \DashLine(130,5)(130,120){6}
  \put(  3, 68){$f$}
  \put( 85, 75){$f$}
  \put( 72, 60){$\gamma$}
  \put( 40,  2){$a$}
  \put( 25, 85){$p_f$}
  \put( 72,100){$p'_f$}
  \put( 52, 42){$k$}
  \put(108, 36){\huge$\vdots$\normalsize $\,X$}
  \put( 78,115){${\cal M}_{fa\to fX}$}
  \put(140,115){${\cal M}_{fa\to fX}^*$}
  \end{picture} } 
\put(240,-5){ 
  \begin{picture}(210,120)(0,0)
  \Photon(90,40)(50,70){2}{5}
  \Photon(210,70)(170,40){2}{5}
  \Line(90,40)( 50, 10)
  \Line(170,40)(210, 10)
  \LongArrow( 55,57)( 70, 46)
  \CArc(130,10)(50,45,135)
  \CArc(130,70)(50,-135,-45)
  \GCirc( 90,40){10}{1}
  \GCirc(170,40){10}{1}
  \DashLine(130,5)(130,95){6}
  \put( 72, 60){$\gamma$}
  \put( 40,  2){$a$}
  \put( 52, 42){$k$}
  \put(108, 36){\huge$\vdots$\normalsize $\,X$}
  \put( 82,90){${\cal M}_{\gamma a\to X}$}
  \put(140,90){${\cal M}_{\gamma a\to X}^*$}
  \end{picture} } 
\end{picture} }
\caption{Generic squared diagram for the splitting $f\to f\ga^*$ 
(left) and the corresponding squared diagram for the related process
with an incoming $\gamma$ (right).}
\label{fig:ffast2}
\efi
Assuming summation over the polarization of the outgoing fermion $f$,
the squared matrix element behaves like
\beqar
\lefteqn{
|\M_{fa\to fX}(p_f,p_a,p'_f;\kappa_f)|^2}
\nn\\
&\asymp{p_f p'_f\to0}&
\frac{N_{\mathrm{c},f} Q_f^2 e^2}{k^4} \, 
\mathrm{Tr}\left\{ \Sigma_{\kappa_f} (\dsl{p}_f+m_f) \gamma_\mu
(\dsl{p}'_f+m_f) \gamma_\nu \right\} \, 
T_{\gamma a\to X}^\mu(k,p_a)^* \,
T_{\gamma a\to X}^\nu(k,p_a),
\hspace{2em}
\label{eq:M2ffast}
\eeqar
where $T_{\gamma a\to X}^\mu(k,p_a)$ includes all information of the 
subamplitude indicated by the open blob in \reffi{fig:ffast2}.
To leading order in $m_f\to0$,
the squared amplitude for the subprocess $\gamma a\to X$ can
be written as
\beq
|\M_{\ga a\to X}(\tilde k,p_a;\la_\ga)|^2 = 
\veps_{\la_\ga,\mu}^*(\tilde k)
T_{\gamma a\to X}^\mu(\tilde k,p_a)^* \,
\veps_{\la_\ga,\nu}(\tilde k) \,
T_{\gamma a\to X}^\nu(\tilde k,p_a),
\eeq
with the helicity $\la_\ga=\pm$ of the incoming photon and
the light-like vector $\tilde k^\mu$.
In order to relate the $fa$ process with the $\gamma a$ subprocess,
we first evaluate the trace in Eq.~\refeq{eq:M2ffast} and drop
all terms that vanish owing to the Ward identity
$k_\mu T_{\gamma a\to X}^\mu(k,p_a)=0$.
Inserting the form \refeq{eq:spfkn} of $s^\mu_{p_f}$,
the result can be written as
\beqar
\lefteqn{
|\M_{fa\to fX}(p_f,p_a,p'_f;\kappa_f)|^2}
\nn\\
&\asymp{p_f p'_f\to0}&
\frac{N_{\mathrm{c},f} Q_f^2 e^2}{-k^2} \, 
\biggl\{ -g_{\mu\nu} - \frac{4p_{f,\mu}p_{f,\nu}}{k^2}
-\frac{\ri\kappa_f}{k^2} \,\eps_{\mu\nu\alpha\beta}\, k^\alpha
\biggl( \frac{(p_f n)}{k_0} \tilde k^\beta 
-\frac{m_f^2}{(p_f n)} n^\beta \biggr)
\biggr\} 
\nn\\
&& {} \times
T_{\gamma a\to X}^\mu(k,p_a)^* \,
T_{\gamma a\to X}^\nu(k,p_a).
\hspace{2em}
\label{eq:M2ffast2}
\eeqar
Now we make use of the collinear limit which is characterized by
$p_f p'_f=m_f^2-p_f k={\cal O}(m_f^2)\to0$.
We decompose the photon momentum according to
\beq
k^\mu = xp_f^\mu + k_\perp^\mu + k_r^\mu
\label{eq:kparam}
\eeq
with $x=k^0/p_f^0$, $p_fk_\perp=0$, and ${\bf k}_r={\bf 0}$.
In this decomposition we have 
${\cal O}(k_\perp^0)={\cal O}(k_r^0)= {\cal O}(m_f^2)$ and
${\bf k}_\perp^2={\cal O}(m_f^2)$, i.e.\
the vector $k^\mu_\perp$ can be counted as ${\cal O}(m_f)$.
Moreover, we get $k^2_\perp=x^2m_f^2+k^2(1-x)+{\cal O}(m_f^4)$.
In the determination of the leading collinear behaviour of 
Eq.~\refeq{eq:M2ffast2}, we can replace the momentum $k^\mu$
by the light-like momentum $\tilde k^\mu=k^\mu+{\cal O}(m_f)$
in the two $T_{\gamma a\to X}(k,p_a)$ terms.
The expansion of the two terms with the $\eps$-tensor
is also straightforward. With the help of identity~\refeq{eq:apolid},
the contraction $\eps_{\mu\nu\alpha\beta} k^\alpha n^\beta$ 
becomes
\beq
\eps_{\mu\nu\alpha\beta} k^\alpha n^\beta = 
\eps_{\mu\nu\alpha\beta} \tilde k^\alpha n^\beta +{\cal O}(m_f) =
\ri(kn)\left[\veps_{+,\mu}(\tilde k)\veps_{-,\nu}(\tilde k)
-\veps_{-,\mu}(\tilde k)\veps_{+,\nu}(\tilde k)\right] +{\cal O}(m_f).
\label{eq:eps1}
\eeq
The second contraction $\eps_{\mu\nu\alpha\beta} k^\alpha \tilde k^\beta$
can be expanded upon writing
$\eps_{\mu\nu\alpha\beta}= {g_\mu}^{\mu'}{g_\nu}^{\nu'}{g_\alpha}^{\alpha'}
\eps_{\mu'\nu'\alpha'\beta}$
with the following decomposition of the metric tensor,
\beq
g^{\mu\nu} = \frac{1}{2k_0}(n^\mu \tilde k^\nu+\tilde k^\mu n^\nu)
-\veps_+^\mu(\tilde k)\veps_-^\nu(\tilde k)
-\veps_-^\mu(\tilde k)\veps_+^\nu(\tilde k).
\label{eq:metricdecomp}
\eeq
The $\eps$-tensor now only appears as 
$\eps_{\mu\nu\alpha\beta}\veps_+^\mu(\tilde k)\veps_-^\nu(\tilde k)
n^\alpha \tilde k^\beta = 2\ri k_0$, and the momentum $\tilde k^\mu$
with an open index can be replaced via
\beq
\tilde k^\mu = \frac{2k_0}{(kn)}
\left[ k^\mu +\left(\veps_-(\tilde k)\cdot k\right)\veps_+^\mu(\tilde k)
+\left(\veps_+(\tilde k)\cdot k\right)\veps_-^\mu(\tilde k) \right]
+{\cal O}(m_f^2),
\eeq
which follows from Eq.~\refeq{eq:metricdecomp} upon contraction with
$k^\nu$. This procedure spans the tensor 
$\eps_{\mu\nu\alpha\beta} k^\alpha \tilde k^\beta$ in terms of 
$\veps_{\pm,\mu}(\tilde k)\veps_{\mp,\nu}(\tilde k)$ and covariants
involving $k_\mu$ or $k_\nu$. The latter do not contribute because
of the Ward identity $k_\mu T^\mu_{\gamma a\to X}(k,p_a)=0.$
The expansion of the various scalar products for $m_f\to0$ 
is straightforward, yielding
\beqar
\eps_{\mu\nu\alpha\beta} k^\alpha \tilde k^\beta &=&
\frac{\ri k_0[k^2(x-2)-x^2 m_f^2]}{x(p_f n)}
\left[\veps_{+,\mu}(\tilde k)\veps_{-,\nu}(\tilde k)
-\veps_{-,\mu}(\tilde k)\veps_{+,\nu}(\tilde k)\right] 
\nn\\
&& {}
+ (\mbox{terms proportional to $k_\mu$ or $k_\nu$})
+ {\cal O}(m_f^3).
\label{eq:eps2}
\eeqar
Inserting Eqs.~\refeq{eq:eps1} and \refeq{eq:eps2} into
Eq.~\refeq{eq:M2ffast2} and performing the expansion
in the collinear limit leads to the form given in 
Eqs.~\refeq{eq:ffast-fact} and \refeq{eq:hff}.
\end{sloppypar}

The final step of performing the azimuthal average around the
collinear axis, which leads to Eqs.~\refeq{eq:ffast-fact2} and
\refeq{eq:hff2}, is most easily carried out by fixing a specific
coordinate frame. In a frame where the direction of ${\bf p}_f$
is given by ${\bf e}_f^{\mathrm{T}}=(0,0,1)$, the vectors
$\tilde k^\mu$ and $\veps^\mu_\pm(\tilde k)$ are given by
\beq
\tilde k^\mu = (k_0,0,0,k_0), \qquad
\veps^\mu_\pm(\tilde k) = \frac{1}{\sqrt{2}}(0,1,\pm\ri,0).
\eeq
Recall that $k^\mu$ and $\tilde k^\mu$ differ only by terms of
${\cal O}(m_f)$ in the collinear limit. According to definition
\refeq{eq:kparam} the leading term of $k^\mu_\perp$ takes the form
\beq
k^\mu_\perp = (0,-|{\bf k}_\perp|\cos\phi'_f,-|{\bf k}_\perp|\sin\phi'_f,0)
+ {\cal O}(m_f^2),
\eeq
where $\phi'_f$ is the azimuthal angle of ${\bf p}'_f={\bf p}_f-{\bf k}$.
In this parametrization the average 
$\langle k^\mu_\perp k^\nu_\perp\rangle_{\phi'_f}$ is easily calculated to
\beq
\langle k^\mu_\perp k^\nu_\perp\rangle_{\phi'_f} 
= -\frac{k^2_\perp}{2} \mathrm{diag}(0,1,1,0) + {\cal O}(m_f^3)
= -\frac{k^2_\perp}{2} E^{\mu\nu}(\tilde k) + {\cal O}(m_f^3),
\eeq
while it is trivially seen that the tensors
$\veps^\mu_\pm(\tilde k) \veps^\nu_\pm(\tilde k)^*$ 
do not change after taking the azimuthal average.
With these considerations the transition from 
Eqs.~\refeq{eq:ffast-fact} and \refeq{eq:hff} to
the averaged form in Eqs.~\refeq{eq:ffast-fact2} and
\refeq{eq:hff2} is straightforward.

\subsection{\boldmath{Dipole subtraction for $f\to f\ga^*$ splittings
with massive final-state spectator}}
\label{app:ffast-fin}

In \refse{se:ffast-fin} we have presented all formulas needed for
the case of a massless final-state spectator in practice, but
did not go into the details of their derivation. Here we close this
gap by deriving the formalism in the more general situation of a
possibly massive spectator $j$.
We keep the general definition \refeq{eq:Msubffast-fin} of the
subtraction function, but generalize the subtraction kernel
as follows,
\beqar
h_{j,\kappa_f}^{ff,\mu\nu}(p_f,p'_f,p_j) &=& 
\frac{-1}{(p_f-p'_f)^2} \Biggl[-g^{\mu\nu}-\frac{4(1-x)}{x^2}
\frac{\tilde k_\perp^\mu\tilde k_\perp^\nu}{\tilde k_\perp^2} \,
\frac{(p_f-p'_f)^2(1-x)+m_f^2 x^2}{(p_f-p'_f)^2(1-x)}
\nn\\
&& \hspace{3em} {}
+ \frac{\kappa_f}{x}\Biggl(2-x+\frac{2x^2 m_f^2}{(p_f-p'_f)^2}\Biggr)
\Bigl( \veps^\mu_+(\tilde k)^*\veps^{\nu}_+(\tilde k)
-\veps^\mu_-(\tilde k)^*\veps^{\nu}_-(\tilde k) \Bigr)
\Biggr],
\nn\\
\eeqar
because we need the correct dependence on the emitter mass $m_f$
for the integration of $|\M_{\sub}|^2$ below.
The auxiliary parameters still have the form \refeq{eq:xzffast},
but the auxiliary momenta become more complicated,
\beqar
\tilde k^\mu(x) &=& \frac{m_j^2-P^2}{-\bar P^2} \frac{x}{R(x)}
\biggl( p_f^\mu + \frac{\bar P^2+2m_f^2 x}{2xP^2} P^\mu \biggr)
+\frac{m_j^2-P^2}{2P^2} P^\mu, \qquad
\tilde k^\mu = \tilde k^\mu(x_{fj,f}),
\nn\\
\tilde p_j(x) &=& P^\mu+\tilde k^\mu(x), \qquad
\tilde p_j = \tilde p_j(x_{fj,f}),
\nn\\
\tilde k_\perp^\mu &=& 
 \frac{p_j\tilde k}{\tilde p_j\tilde k} \, p_f^{\prime\mu}
-\frac{p'_f\tilde k}{\tilde p_j\tilde k} \, p_j^\mu.
\eeqar
Here we made use of the abbreviations
\beq
P^\mu=p^{\prime\mu}_f+p^\mu_j-p^\mu_f, \qquad
\bar P^2 = P^2-2m_f^2-m_j^2, \qquad
R(x) = \frac{\sqrt{(\bar P^2+2m_f^2 x)^2-4x^2 m_f^2 P^2}}{-\bar P^2},
\eeq
The new momenta satisfy the on-shell conditions $\tilde k^2=0$,
$\tilde p_j^2=m_j^2$ and correctly behave in the collinear limit,
$\tilde k\to x p_f$, where $p_f p'_f={\cal O}(m_f^2)\to0$.
Before carrying out the singular integrations, we average the
subtraction function over $\phi'_f$, yielding
\beq
\langle |\M_{\sub}(\kappa_f)|^2 \rangle_{\phi'_f} =
N_{\mathrm{c},f}\,Q_f^2e^2 \, 
h^{ff}_{j,\tau}(p_f,p'_f,p_j) \,
|\M_{\ga a\to jX}(\tilde k,p_a;\la_\ga=\tau\kappa_f)|^2
\eeq 
with summation over $\tau=\pm$ and
\beqar
h^{ff}_{j,\tau}(p_f,p'_f,p_j) &=& 
\frac{-1}{\bar P^2 z_{fj,f}+2m_f^2 x_{fj,f}}
\Biggl[
P_{\ga f}(x_{fj,f})
+ \frac{2m_f^2x_{fj,f}^2}{\bar P^2 z_{fj,f}+2m_f^2 x_{fj,f}}
\nn\\
&& \phantom{\frac{-1}{\bar P^2 z_{fj,f}+2m_f^2 x_{fj,f}} \Biggl[} {}
+\tau\left(2-x_{fj,f}+\frac{2x_{fj,f}^3 m_f^2}
{\bar P^2 z_{fj,f}+2m_f^2 x_{fj,f}}\right)
\Biggr].
\hspace{2em}
\eeqar

The splitting of the $(N+1)$-particle phase space into the corresponding
$N$-particle phase space and the integral over the remaining
singular degrees of freedom is given by
\beq
\int\rd\phi(p'_f,p_j,k_X;p_f+p_a) 
= \int_{0}^{x_1}\rd x
\int\rd\phi\Big(\tilde p_j(x),k_X;\tilde k(x)+p_a\Big)
\int [\rd p'_f(P^2,x,z_{fj,f})],
\eeq
with the explicit parametrization 
\beq
\int [\rd p'_f(P^2,x,z_{fj,f})] = \frac{1}{4(2\pi)^3} 
\frac{\tilde s}{\bar s} \, \frac{-\bar P^2}{x^2 R(x)} \,
\int_{z_1(x)}^{z_2(x)} \rd z_{fj,f} \,
\int \rd\phi'_f.
\eeq
The upper kinematical limit of the parameter $x=x_{fj,f}$ is given by
\beq
x_1 = \frac{-\bar P^2}{-\bar P^2+2m_f m_j}.
\eeq
The integration of the azimuthal angle $\phi'_f$ of $f(p'_f)$ simply
yields a factor $2\pi$. The non-trivial integration over $z_{fj,f}$
has the boundary
\beq
z_{1,2}(x) = \frac{2m_f^2 x+\bar P^2(x-1) \mp R(x)
	\sqrt{\bar P^4(1-x)^2-4m_f^2 m_j^2 x^2}}%
		{2[\bar P^2(x-1)+(m_f^2+m_j^2)x]}.
\eeq

Defining the integrated subtraction kernel according to
\beq
\cHsub^{ff}_{j,\tau}(P^2,x) = \frac{-\bar P^2}{2R(x)}
\int_{z_1(x)}^{z_2(x)} \rd z_{fj,f} \, h^{ff}_{j,\tau}(p_f,p'_f,p_j),
\eeq
the cross-section contribution of the subtraction part takes
the form \refeq{eq:sigffast-fin} in the limit $m_f\to0$.
For a non-zero spectator mass $m_j$, the function $\cHsub^{ff}_{j,\tau}(P^2,x)$
reads
\beq
\cHsub^{ff}_{j,\tau}(P^2,x) = \frac{1}{2}
\ln\biggl(\frac{\bar P^4(1-x)^2}{x^3 m_f^2[-\bar P^2(1-x)+m_j^2 x]}\biggr)
\biggl[ P_{\ga f}(x) + \tau (2-x)\biggr] - \frac{1-x}{x} -\tau(1-x).
\eeq

\end{document}